	\documentclass{sn-jnl}

\usepackage{manyfoot}
\usepackage{xcolor}

\usepackage[square,numbers]{natbib}

\usepackage{amsmath,amsthm,amssymb}
\usepackage{url}
\usepackage{bm}
\usepackage{graphicx}  
\usepackage{tikz}
\usetikzlibrary{calc}

\usepackage{enumerate}
\usepackage[mathscr]{euscript}

\begin{document}

\newcommand{\IGNORE}[1]{}

\title[Generalizing the WGMV Primal-Dual Method]{Improved Approximation Algorithms by Generalizing the Primal-Dual Method Beyond Uncrossable Functions}

\author[1]{\fnm{Ishan} \sur{Bansal}}\email{ib332@cornell.edu} 
\author*[2]{\fnm{Joseph} \sur{Cheriyan}}\email{jcheriyan@uwaterloo.ca} 
\author[1]{\fnm{Logan} \sur{Grout}}\email{lcg58@cornell.edu} 
\author[3]{\fnm{Sharat} \sur{Ibrahimpur}}\email{S.Ibrahimpur@lse.ac.uk}

\affil[1]{Cornell University, Operations Research and Information Engineering, Ithaca, NY, USA}
\affil[2]{Department of Combinatorics and Optimization, University of Waterloo, Canada}
\affil[3]{Department of Mathematics, London School of Economics and Political Science, UK}


\setlength{\topsep}{1em} 
\newtheorem{theorem}{Theorem}[section]
\newtheorem{proposition}[theorem]{Proposition}
\newtheorem{lemma}[theorem]{Lemma}
\newtheorem{fact}[theorem]{Fact}
\newtheorem{claim}[theorem]{Claim}
\newtheorem{definition}{Definition}[section]
\newtheorem{example}{Example}[section]




\newenvironment{proofof}[1]{\begin{proof}[{Proof of #1}]}{\end{proof}}

\newcommand{\R}{\ensuremath{\mathbb R}}
\newcommand{\Rp}{\ensuremath{\R_{\geq 0}}}
\newcommand{\Q}{\ensuremath{\mathbb Q}}
\newcommand{\Qp}{\ensuremath{\Q_{\geq 0}}}
\newcommand{\Zint}{\ensuremath{\mathbb Z}}
\newcommand{\Zp}{\ensuremath{\Zint_{\geq 0}}}

\newcommand{\opt}{\textsc{opt}}
\newcommand{\safe}{\mathscr{S}}
\newcommand{\unsafe}{\mathscr{U}}

\newcommand{\req}{\textsc{req}}

\newcommand{\cp}{\ensuremath{\widehat{\cal P}}}

\newcommand{\fgc}{\mathrm{FGC}}
\newcommand{\pqfgc}{(p,q)\text{-}\fgc}
\newcommand{\ponefgc}{(p,1)\text{-}\fgc}
\newcommand{\ptwofgc}{(p,2)\text{-}\fgc}
\newcommand{\pthreefgc}{(p,3)\text{-}\fgc}
\newcommand{\oneonefgc}{(1,1)\text{-}\fgc}
\newcommand{\oneqfgc}{(1,q)\text{-}\fgc}
\newcommand{\pfourfgc}{(p,4)\text{-}\fgc}
\newcommand{\twopfourfgc}{(2p,4)\text{-}\fgc}

\newcommand{\As}{\mathscr{A}} 
\newcommand{\C}{\mathscr{C}} 
\newcommand{\F}{\mathcal{F}} 
\newcommand{\Ls}{\mathscr{L}} 
\newcommand{\Ps}{\mathscr{P}} 
\newcommand{\Ss}{\mathscr{S}} 
\newcommand{\T}{\mathscr{T}} 
\newcommand{\mcS}{\Ss}
\newcommand{\ba}{\ensuremath{\mathbbmss{a}}}

\newcommand{\tc}{\tilde{c}}
\newcommand{\tG}{\widetilde{G}}
\newcommand{\tN}{\tilde{N}}
\newcommand{\tV}{\widetilde{V}}
\newcommand{\tQ}{\widetilde{Q}}
\newcommand{\tE}{\widetilde{E}}

\newcommand{\hF}{\hat{F}}
\newcommand{\hH}{\hat{H}}

\newcommand{\sm}{\setminus}
\newcommand{\wt}{\widetilde}

\newcommand{\myredtext}[1]{#1}
\newcommand{\redtext}[1]{{\color{red}{#1}}}
\newcommand{\bluetext}[1]{{\color{blue}{#1}}}
\newcommand{\greentext}[1]{{\color{green}{#1}}}
\newcommand{\unfn}{f}
\newcommand{\genuntext}{pliable}
\newcommand{\genun}{pliable\ }
\newcommand{\GenUn}{Pliable\ }
\newcommand{\genunnogap}{pliable}
\newcommand{\propred}{\ensuremath{(\gamma)}}
\newcommand{\capbound}{\wt{\lambda}}
\newcommand{\augsmallcuts}{\mathrm{AugSmallCuts}}

\newcommand{\genfgc}[2]{\ensuremath{(#1,#2)}\text{-}\fgc}
\newcommand{\capndp}{\mathrm{Cap\text{-}NDP}}

\newcommand{\pliableapx}{16}
\newcommand{\oddapx}{20}
\newcommand{\nearmincutsapx}{O(1)}
\newcommand{\ptwofgcapx}{O(1)}
\newcommand{\primaldualpliableapx}{O(1)}


\newcommand{\cross}{\bowtie}
\newcommand\ria{\rightarrow}

\newcommand{\fconn}{\ensuremath{f\text{-}\mathrm{connectivity}}}

\newcommand{\fcproblem}{\ensuremath{f\text{-}\mathrm{connectivity~problem}}}

\abstract{
We address long-standing open questions raised by Williamson,
Goemans, Vazirani and Mihail pertaining to the design of approximation
algorithms for problems in network design via the primal-dual method
(Combinatorica 15(3):435-454, 1995).
Williamson et~al.\ prove an approximation ratio of two for
connectivity augmentation problems where the connectivity requirements
can be specified by uncrossable functions.
They state: ``Extending our algorithm to handle non-uncrossable
functions remains a challenging open problem.
The key feature of uncrossable functions is that there exists an
optimal dual solution which is laminar~\dots\ A larger
open issue is to explore further the power of the primal-dual
approach for obtaining approximation algorithms for other combinatorial
optimization problems.''

\vspace{0.40em}

Our main result proves that the primal-dual algorithm of Williamson et~al.\ 
achieves an approximation ratio of  $\pliableapx$
for a class of functions that generalizes the notion of an uncrossable function.
There exist instances that can be handled by our methods where none of the optimal dual solutions has a laminar support.

\vspace{0.40em}

We present three applications of our main result to problems in the area of network~design.


\vspace{0.225em}

(1)~ A $\pliableapx$-approximation algorithm for augmenting a family of
small cuts of a graph $G$.
The previous best approximation ratio was $O(\log{|V(G)|})$.

\vspace{0.225em}

(2)~ A $\pliableapx \cdot {\lceil k/u_{min} \rceil}$-approximation
algorithm for the Cap-$k$-ECSS problem which is as follows:
Given an undirected graph $G = (V,E)$ with edge costs $c \in \Qp^E$ and
edge capacities $u \in \mathbb{Z}_{\geq 0}^E$, find a minimum-cost
subset of the edges $F\subseteq E$ such that the capacity of any
cut in $(V,F)$ is at least $k$;
$u_{min}$ (respectively, $u_{max}$) denotes the minimum (respectively,
maximum) capacity of an edge in $E$, and w.l.o.g.\ $u_{max} \leq k$.
The previous best approximation ratio was $\min(O(\log{|V|}), k, 2u_{max})$.

\vspace{0.225em}

(3)~ A $\oddapx$-approximation algorithm for the model of $(p,2)$-Flexible Graph Connectivity.
The previous best approximation ratio was $O(\log{|V(G)|})$, where $G$ denotes the input graph.
}

\maketitle

{
\section{Introduction} \label{sec:intro}
The primal-dual method is a well-known algorithmic discovery of the past century.
Kuhn (1955) \cite{Kuhn} presented a primal-dual algorithm for weighted bipartite matching,
and Dantzig et al.\ (1957) \cite{DFF} presented a generalization for solving linear programs.
Primal-dual methods for problems in combinatorial optimization are based
on linear programming (LP) relaxations; the associated linear programs
(LPs) are crucial for the design and analysis of these algorithms.
A key feature of the primal-dual method is that it leads to self-contained
combinatorial algorithms that do not require solving the underlying LPs.
This key feature makes it attractive for both theoretical studies and
real-world computations.

Several decades after the pioneering work of Kuhn, the design of
approximation algorithms for NP-hard problems emerged as an important
area of research.
Agrawal, Klein and Ravi \cite{AKR95} designed and analyzed a primal-dual
approximation algorithm for the Steiner forest problem.
Goemans and Williamson \cite{GW95} generalized  this algorithm to
constrained forest problems.  Subsequently, Williamson, Goemans,
Vazirani and Mihail \cite{WGMV95} (abbreviated WGMV)  extended the methods
of \cite{GW95} to obtain a primal-dual $2$-approximation algorithm for the
problem of augmenting the connectivity of a graph to satisfy requirements
specified by {uncrossable} functions.  These functions are versatile
tools for modeling several key problems in the area of network~design.

Network design encompasses diverse problems that find applications
in sectors like transportation, facility location, information security,
and resource connectivity, to name a few.
The area has been studied for decades and it has led to major algorithmic
innovations. Most network-design problems are NP-Hard, and many of these
problems are APX-hard.
Consequently, research in the area has focused on the design and analysis
of good approximation algorithms, preferably with a small constant-factor
approximation ratio.
In the context of network design, many of the $O(1)$~approximation
algorithms rely on a particular property called \emph{uncrossability},
see the books by Lau, Ravi \& Singh \cite{LRS-book}, Vazirani
\cite{Vazirani-book}, and Williamson \& Shmoys \cite{WS-book}.
This property has been leveraged in various ways to obtain algorithms
with $O(1)$~approximation ratios for problems such as survivable network
design \cite{Jain01}, min-cost/min-size $k$-edge connected spanning
subgraph \cite{GGTW09,GabowGallagher12}, min-cost 2-node connected
spanning subgraph \cite{FJW06}, $(p,1)$-flexible graph connectivity
\cite{BCHI23}, etc. The primal-dual method is one of the most successful
algorithmic paradigms that leverages these uncrossability properties.
On the other hand, when the uncrossability property does not hold, then
most of the known techniques in this area for designing $O(1)$~approximation algorithms
fail to work.
WGMV~\cite{WGMV95} conclude their paper with the following remark:
\begin{quote}
Extending our algorithm to handle non-uncrossable functions remains a challenging open problem.
The key feature of uncrossable functions is that there exists an
optimal dual solution which is laminar~\dots\ A larger
open issue is to explore further the power of the primal-dual
approach for obtaining approximation algorithms for other combinatorial
optimization problems.
Handling all non-uncrossable functions is ruled out by the fact
that there exist instances corresponding to non-uncrossable $\{0,1\}$
functions whose relative duality gap is larger than any constant.
\end{quote}

Our main contribution is an extension of the WGMV primal-dual algorithm
and its analysis to a class of functions that is more general than
the class of uncrossable functions.
Our main result is an approximation ratio of $\pliableapx$ for the larger class of functions.
We apply our main result to give improved approximation ratios
for three problems:
(i)~augmenting all small cuts of a graph,
(ii)~finding a minimum-cost capacitated $k$-edge connected subgraph, and
(iii)~the $\ptwofgc$ problem.
We give a detailed discussion of our results in
Sections~\ref{sec:intro:mainresult}--\ref{sec:intro:p2fgc}.

Primal-dual algorithms for solving network design problems work iteratively,
starting with a graph that has no edges,
and adding a set of edges in each iteration,
until the set of picked edges forms a feasible subgraph
(i.e., a subgraph satisfying the requirements of the problem).
In our context, the problem requirements are formalized by specifying
a connectivity requirement (which is a non-negative integer) for each
set of nodes.
Consider one of the iterations.
Let $F$ denote the set of edges picked until the start of this iteration.
A set of nodes $S$ is said to be \emph{violated} if the number of
$F$-edges in the cut of $S$ is less than the pre-specified connectivity
requirement of $S$.
An edge is deemed to be useful if it is in the cut of a violated set $S$.
The goal of the iteration is to buy a cheap set of useful edges.

Clearly, the family of violated sets is important for the design and analysis of these algorithms.
(Recall that sets $A,B \subseteq V$ are said to \emph{cross} if each
of the four sets $A\cap{B}$, $V\sm(A\cup{B})$, $A\sm{B}$, $B\sm{A}$ is nonempty.)
A family $\F$ of sets is called \emph{uncrossable} if the following holds: 
\[
A,B \in \F \: \implies \: 
	A\cap{B}, A\cup{B}\in\F \text{ or } A\sm{B},B\sm{A}\in\F.
\]
Informally speaking, the uncrossability property of $\F$ ensures that
the (inclusion-wise) minimal sets of $\F$ can be considered independently.
Now, suppose that $\F$ is the family of violated sets at some iteration
of the primal-dual algorithm, and suppose that $\F$ is uncrossable.
Then observe that an (inclusion-wise) minimal violated set $A\in\F$
cannot cross another set $S\in\F$; otherwise, we get a contradiction
since $A,S\in\F$ implies that a proper subset of $A$ is in $\F$
(either $A\cap{S}\in\F$ or $A\sm{S}\in\F$, because $\F$ is uncrossable).
This key property is one of the levers used in the design of some of
the $O(1)$-approximation algorithms for problems in network~design.

Unfortunately, there are important problems in network
design where the family of violated sets does not form an
uncrossable family; for instance, see the discussion in Appendix~\ref{append:nonlaminar}.
Let us call a family of sets $\F$ \emph{pliable} if the following holds:
\[ A,B\in\F \:\implies\: \text{at least two of } A\cap{B}, A\cup{B}, A\sm{B},B\sm{A} \text{ are in } \F.
\]
Clearly, an uncrossable family of sets is a pliable family.
In Appendix~\ref{append:cylinders}, we show that the WGMV primal-dual
algorithm has a super-constant approximation ratio for \genun families.
In more detail, we present an example with $n$ nodes such that the cost
of the solution found by the WGMV algorithm is $\Omega(\sqrt{n})$ times
the cost of an optimal solution.
Nevertheless, we prove an $O(1)$ approximation ratio for the
WGMV primal-dual algorithm applied to \genun families satisfying
an additional property that we call property~\propred;
the formal definition is given below.
Property~\propred\ allows a minimal violated set to cross another violated
set, but, crucially, it does not allow a minimal violated set to cross
an arbitrary number of violated sets in arbitrary ways.
Thus, \genun families satisfying property~\propred\ are (strictly) more
general than uncrossable families.
Informally speaking, this is what allows us to obtain improved approximation ratios
for problems such as $\ptwofgc$ via the WGMV primal-dual algorithm.
Now, we state property~\propred.

\begin{minipage}{\textwidth}{
\begin{align*}
\hspace{-11pt}
\textbf{Property~\boldmath\propred}: \, 
&\text{For any $F\subseteq{E}$ and for any violated sets (w.r.t.~$F$) $C, S_1, S_2,$ such that} \\
&	(i)~S_1\subsetneq S_2,
	(ii)~C \text{ is inclusion-wise minimal}, \text{ and }
	(iii)~C \text{ crosses      both } S_1, S_2,	\\
&\qquad \qquad \text{$S_2 \setminus (S_1\cup C)$ is either empty or violated.} 
\end{align*}
}\end{minipage}

{
\begin{figure}[htb] \centering
{
\begin{tikzpicture} 
\begin{scope}[every node/.style={circle, fill=black, draw, inner sep=0pt, minimum size = 0.15cm, label distance=0.1cm}]
	\node[label=below:$v_6$] (e) at (0,0) {};
	\node[label={[label distance=0.0]90:$v_1$}]	(a) at ($(e)+(-3,1.5)$) {};
	\node[label={[label distance=0.0]90:$v_2$}]	(ab) at ($(e)+(-2,1.5)$) {};
	\node[label={[label distance=0.0]90:$v_3$}]	(b) at ($(e)+(-1,1.5)$) {};
	\node[label={[label distance=0.0]90:$v_4$}]	(c) at ($(e)+(1,1.5)$) {};
	\node[label={[label distance=0.0]90:$v_5$}]	(d) at ($(e)+(3,1.5)$) {};
\end{scope}

\begin{scope}[line width=0.6pt]
	\draw[thick] ($0.5*(a)+0.5*(b)$) ellipse (1.5cm and 0.75cm);
	\node at ($0.75*(a)+0.25*(b)+(0,1.0)$) {$C$};
	
	\draw[thick,dashed] ($0.5*(b)+0.5*(c)$) ellipse (1.5cm and 0.75cm);
	\node at ($0.25*(b)+0.75*(c)+(0,1.0)$) {$S_1$};
	
	\draw[thick,dashed] ($0.5*(ab)+0.5*(d)$) ellipse (3.00cm and 1.25cm);
	\node at ($0.25*(b)+0.75*(d)+(0,1.4)$) {$S_2$};
\end{scope}
\end{tikzpicture}
}
\caption{
        \label{f:property-propred}
        Illustration of property~\propred, and the sets $C, S_1, S_2$.
	The set $S_2 \setminus (S_1\cup C)$ is non-empty.
}               
\end{figure}
}               

\subsection{$f$-connectivity~problem} \label{sec:intro:f-connectivity}

Most connectivity augmentation problems can be formulated in
a general framework called $f$-connectivity. In this problem, we
are given an undirected graph $G = (V,E)$ on $n$ nodes with
nonnegative costs $c \in \Qp^E$ on the edges and a requirement
function $f:2^V\to\{0,1\}$ on subsets of nodes.
We are interested in finding an edge-set $F \subseteq E$ with minimum
cost $c(F) := \sum_{e \in F} c_e$ such that for all cuts
$\delta(S),\ S \subseteq V$, we have
$|\delta(S) \cap F| \geq f(S)$.
This problem can be formulated as the following integer program
where the binary variable $x_e$ models the inclusion of the edge $e$ in $F$:

\begin{align*} \tag{$f$-IP}
\label{eq:fconnectivityIP}
\min \quad \qquad		& \quad \sum_{e \in E} c_e x_e 		& \\
\text{subject to: } 	& \quad x( \delta(S) ) \geq f(S) 	& \forall \, \, S \subseteq V \\
					& \quad x_e \in \{0,1\} & \forall \, \, e \in E. \\
\end{align*}

In its most~general form, the $\fcproblem$ is hard to approximate
within a logarithmic factor. This can be shown via a reduction from
the hitting set problem which has a logarithmic hardness-of-approximation
threshold; see the discussion at the end of Section~\ref{sec:prelims-graphs}.

If our goal is to obtain $O(1)$-approximation algorithms,
then we must require $f$ to have some properties such that the
corresponding $\fcproblem$ captures some interesting problems in
connectivity augmentation, while ensuring that the $\fcproblem$
does not capture problems that have super-constant
hardness-of-approximation thresholds.
This motivates the following definitions.

\begin{definition}[\cite{WGMV95}] \label{def:uncrossable}
A function $f : 2^V \to \{0,1\}$ satisfying $f(V) = 0$
is called \emph{uncrossable} if for any $A, B \subseteq V$ with $f(A) = f(B) = 1$, we have
$f(A \cap B) = f(A \cup B) = 1$ or $f(A \setminus B) = f(B \setminus A) = 1$.
\end{definition}

\begin{definition} \label{def:pliable}
A function $f : 2^V \to \{0,1\}$ satisfying $f(V) = 0$
is called \emph{\genuntext} if for any $A, B \subseteq V$ with $f(A) = f(B) = 1$, we have
$f(A \cap B) + f(A \cup B) + f(A \setminus B) + f(B \setminus A) \geq 2$.
\end{definition}

Clearly, the problem of augmenting an uncrossable (respectively,
pliable) family of violated sets can be formulated as an $f$-connectivity
problem whose requirement function is an uncrossable (respectively,
pliable) function.

\subsection{\boldmath Our main result} \label{sec:intro:mainresult}

Our main result is that the WGMV primal-dual algorithm achieves an
approximation ratio of $O(1)$ for the $\fcproblem$ whenever $f$
is a \genun function satisfying property~\propred.
As discussed above, the analysis of WGMV relies on the property
that for any (inclusion-wise) minimal violated set $C$ and any violated
set $S$, either $C$ is a subset of $S$ or $C$ is disjoint from $S$
(\cite[Lemma~5.1(3)]{WGMV95}).
This property does not hold
{for}
a \genun function satisfying property~\propred; see the
instance described in Appendix~\ref{append:nonlaminar}.
Informally speaking, our analysis of the WGMV primal-dual algorithm
leverages property~\propred\ and proves an approximation ratio of
$O(1)$; recall that property~\propred\ does {not} allow a
minimal violated set to cross an arbitrary number of violated sets
in arbitrary ways.
{(See Section~\ref{sec:prelims} for definitions of terms
such as violated set w.r.t.\ $f, F$.)}

\begin{theorem} \label{thm:pliablemain}
Let $G = (V,E)$ be an undirected graph with nonnegative costs
$c:E\to\Qp$ on its edges, and let $f:2^V\to\{0,1\}$ be a \genun
function satisfying property~\propred. Suppose that there is a
subroutine that, for any given $F \subseteq E$,
computes all minimal violated sets w.r.t.\ $f$ and $F$.
Then, in polynomial time and using a polynomial number of calls to
the subroutine, we can compute a  $\pliableapx$-approximate
solution to the given instance of the $f$-connectivity problem.
\end{theorem}

In the next three sections, we describe problems in the area of
network~design where Theorem~\ref{thm:pliablemain} gives new/improved
approximation algorithms. In each of these applications, we formulate
an $\fcproblem$ where the function $f$ is a \genun function with
property~\propred.

{
\subsection{Application~1: Augmenting a Family of Small Cuts} \label{sec:intro:asc}
Our first application is on finding a minimum-cost augmentation of
a family of small cuts in a graph.
In an instance of the $\augsmallcuts$ problem, we are given an
undirected capacitated graph $G = (V,E)$ with edge-capacities $u
\in \Qp^E$, a set of links $L \subseteq \binom{V}{2}$ with costs
$c \in \Qp^L$, and a threshold $\capbound \in \Qp$.
A subset $F \subseteq L$ of links is said to \emph{augment} a
node-set $S$ if there exists a link $e \in F$ with exactly one
end-node in $S$.
The objective is to find a minimum-cost $F\subseteq{L}$ that augments
all non-empty $S \subsetneq V$ with $u(\delta_E(S)) < \capbound$.

Some special cases of the $\augsmallcuts$ problem have been studied
previously, and, to the best of our knowledge, there is no previous
publication on the general version of this problem.
Let $\lambda(G)$ denote the minimum capacity of a cut of $G$, thus,
$\lambda(G) := \min\{ u(\delta_E(S)) \;:\; \emptyset \subsetneq S \subsetneq V \}$.
Assuming $u$ is integral and $\capbound = \lambda(G) + 1$, we get
the well-known connectivity augmentation problem for which
constant-factor approximation algorithms are known \cite{FJ81,KT93}.
On the other hand, when $\capbound = \infty$, a minimum-cost spanning
tree of $(V,L)$, if one exists, gives an optimal solution to the problem.

Our main result here is an $O(1)$-approximation algorithm for the $\augsmallcuts$ problem that works for any choice of $\capbound$. The proof of the following theorem is given in Section~\ref{sec:nearmincuts}. 

\begin{theorem} \label{thm:nearmincutsaugmentation}
There is a $\pliableapx$-approximation algorithm for the $\augsmallcuts$ problem. 
\end{theorem}

We refer the reader to Benczur \& Goemans \cite{BG08} and the
references therein for results on the representations of the
near-minimum cuts of graphs; they do not study the problem of
augmenting the near-minimum cuts.
}
{
\subsection{\boldmath Application~2: Capacitated {$k$}-Edge-Connected Subgraph Problem} \label{sec:intro:capkecss}
In the capacitated $k$-edge-connected subgraph problem (Cap-$k$-ECSS),
we are given an undirected graph $G = (V,E)$ with edge costs $c \in
\Qp^E$ and edge capacities $u \in \mathbb{Z}_{\geq 0}^E$. The goal is
to find a minimum-cost subset of the edges $F\subseteq E$ such that the
capacity across any cut in $(V,F)$ is at least $k$, i.e.,
$u(\delta_F(S)) \geq k$ for all non-empty sets $S\subsetneq V$.
Let $u_{max}$ and $u_{min}$, respectively,
denote the maximum capacity of an edge in $E$ and
the minimum capacity of an edge in $E$.
We may assume (w.l.o.g.) that $u_{max}\leq k$.

We mention that there are well-known $2$-approximation algorithms
for the special case of the Cap-$k$-ECSS problem with $u_{max}=u_{min}=1$,
which is the problem of finding a minimum-cost $k$-edge connected spanning subgraph.
Khuller \& Vishkin \cite{KV94} presented a combinatorial 2-approximation algorithm and Jain \cite{Jain01} matched this approximation guarantee via the iterative rounding method.

Goemans et~al.~\cite{GoemansGPSTW94} gave a $2k$-approximation
algorithm for the general {Cap-$k$-ECSS} problem.
Chakrabarty et~al.\ \cite{ChakrabartyCKK15} gave a randomized $O(\log |V(G)|)$-approximation algorithm;
note that this approximation guarantee is independent of $k$ but does depend on the size of the underlying graph.
Recently, Boyd et~al.\ \cite{BCHI23}
improved on these results by providing a $\min(k,\,2u_{max})$-approximation algorithm.
We present a $(\pliableapx \cdot {\lceil k/u_{\min} \rceil})$-approximation
algorithm for the Cap-$k$-ECSS problem.
This gives improved approximation ratios when both $u_{\min}$ and
$u_{\max}$ are sufficiently large;
observe that $(\pliableapx \cdot {\lceil k/u_{\min} \rceil})$ is
less than $\min(k,\,2u_{max})$ when $k \geq u_{\max} \geq u_{\min}
\geq 32$ and $u_{\min} \cdot u_{\max} \geq 16k$.

\begin{theorem}\label{thm:CapkECSS}
There is a $\pliableapx \cdot {\lceil k/u_{min} \rceil}$-approximation
algorithm for the Cap-$k$-ECSS problem.
\end{theorem}
The proof of Theorem~\ref{thm:CapkECSS} is given in Section~\ref{sec:capkecss}.

}
{
\subsection{Application~3: \boldmath {$(p,2)$}-Flexible Graph Connectivity} \label{sec:intro:p2fgc}
Adjiashvili, Hommelsheim and M\"uhlenthaler \cite{AHM21}
introduced the model of Flexible Graph Connectivity that we denote by $\fgc$.
Boyd, Cheriyan, Haddadan and Ibrahimpur \cite{BCHI23} introduced a
generalization of $\fgc$.
Let $p \geq 1$ and $q \geq 0$ be integers. In an instance of the
$(p,q)$-Flexible Graph Connectivity problem, denoted $\pqfgc$, we
are given an undirected graph $G = (V,E)$, a partition of $E$ into
a set of safe edges $\safe$ and a set of unsafe edges $\unsafe$,
and nonnegative edge-costs $c \in \Qp^E$.  A subset $F \subseteq
E$ of edges is feasible for the $\pqfgc$ problem if for any set
$F'$ consisting of at most $q$ unsafe edges, the subgraph $(V, F
\setminus F')$ is $p$-edge connected. The objective is to find a
feasible solution $F$ that minimizes $c(F) = \sum_{e \in F} c_e$.
Boyd et al.\ \cite{BCHI23} presented a $4$-approximation algorithm
for $\ponefgc$ based on the WGMV primal-dual method, and a
$(q+1)$-approximation algorithm for $\oneqfgc$; moreover, they gave
an $O(q \log n)$-approximation algorithm for (general) $\pqfgc$.
Concurrently with our work, Chekuri and Jain \cite{CJ23,CJ22} obtained
$O(p)$-approximation algorithms for $\ptwofgc$, $\pthreefgc$ and
$\twopfourfgc$; in particular, for $\ptwofgc$, they prove an
approximation ratio of $(2p+4)$.
Chekuri and Jain \cite{CJ23,CJ22:rounding} have several other results
for problems in the flexible graph connectivity model; they introduce
and study non-uniform fault models such as $(p,q)$-Flex-SNDP.

Our main result here is an $O(1)$-approximation algorithm for the $\ptwofgc$ problem.

\begin{theorem} \label{thm:ptwofgc}
There is a $\oddapx$-approximation algorithm for the $\ptwofgc$ problem. Moreover, for even $p$, the approximation ratio is $6$. 
\end{theorem}

Note that in comparison to \cite{CJ23,CJ22}, Theorem~\ref{thm:ptwofgc}
yields a better approximation ratio when $p>8$ or $p \in \{2,4,6,8\}$.
For $p=1$, the approximation ratio of $3$ from \cite{BCHI23} is
better than the guarantees given by \cite{CJ23,CJ22} and
Theorem~\ref{thm:ptwofgc}. The proof of Theorem~\ref{thm:ptwofgc}
is given in Section~\ref{sec:p2fgc}.

}
\subsection{Related work}

Goemans \& Williamson \cite{GW95} formulated several problems in
network~design as the $\fcproblem$ where $f$ is a proper function.
A symmetric function $f:2^V\to\Zint_{>0}$ with $f(V)=0$ is said to
be \emph{proper} if $f(A \cup B) \leq \max(f(A), f(B))$ for any
pair of disjoint sets $A, B \subseteq V$.

Jain \cite{Jain01} designed the iterative rounding framework for
the setting when $f$ is {weakly supermodular} and presented a
$2$-approximation algorithm.
A function $f$ is said to be \emph{weakly} \emph{supermodular} if
$f(A) + f(B) \leq \max(f(A \cap B) + f(A \cup B),\; f(A\sm{B}) + f(B\sm{A}))$
for any $A, B \subseteq V$.
One can show that proper functions are weakly supermodular. 
We mention that there are examples of uncrossable functions
that are not weakly supermodular, see \cite{BCHI23}.

Although our paper focuses on theory, let us mention that extensive
computational research over decades shows that
{the primal-dual method works well}.
In particular, computational studies of some of the
well-known primal-dual approximation algorithms have been conducted,
and the consensus is that these algorithms work well in practice,
see \cite[Section~4.9]{GW-bookchapter}, \cite{GGW98}, \cite{JMP},
\cite{MSDM96}, \cite{WG96}.

{After the posting of a preliminary version of our paper \cite{BCGI:arxiv},
Bansal \cite{B2023} and Nutov \cite{N2023} have posted related results.}
}
{
\section{Preliminaries} \label{sec:prelims}
This section has definitions and preliminary results.
Our notation and terms are consistent with \cite{Diestel,Schrijver},
and readers are referred to those texts for further information.

For a positive integer $k$, we use $[k]$ to denote the set $\{1,\dots,k\}$.
Sets $A,B \subseteq V$ are said to \textit{cross}, denoted $A\cross B$,
if each of the four sets $A\cap{B}$, $V\sm(A\cup{B})$, $A\sm{B}$, $B\sm{A}$ is non-empty;
on the other hand, if $A,B$ do not cross, then either $A\cup{B}=V$,
or $A,B$ are disjoint, or one of $A,B$ is a subset of the other one.
A family of sets $\Ls \subseteq 2^V$ is said to be \emph{laminar}
if for any two sets $A, B \in \Ls$ either $A$ and $B$ are disjoint
or one of them is a subset of the other one.

We may use abbreviations for some standard terms, e.g.,
we may use ``$\pqfgc$'' as an abbreviation for ``the $\pqfgc$ problem''.
{In some of our discussions, we may use informal phrasing
such as ``we apply the primal-dual method to augment a \genun function.''}

\subsection{Graphs, Subgraphs, and Related Notions} \label{sec:prelims-graphs}

Let $G=(V,E)$ be an undirected multi-graph
(possibly containing parallel edges but no loops)
with non-negative costs $c\in\Rp^{E}$ on the edges.
We take $G$ to be the input graph, and we use $n$ to denote $|V(G)|$.
For a set of edges $F\subseteq E(G)$, $c(F):=\sum_{e\in F}c(e)$,
and for a subgraph $G'$ of $G$, $c(G'):=c(E(G'))$.
For any instance $G$, we use $\opt(G)$ to denote the minimum cost of a feasible subgraph
(i.e., a subgraph that satisfies the requirements of the problem).
When there is no danger of ambiguity, we use $\opt$ rather than $\opt(G)$.

Let $G=(V,E)$ be any multi-graph, let $A,B\subseteq{V}$ be two disjoint node-sets,
and let
{$F\subseteq{E}$ be an edge-set}.
We denote the multi-set of edges of $G$ with exactly one end-node in each
of $A$ and $B$ by $E(A,B)$, thus, $E(A,B) := \{ e=uv : u\in A, v\in B \}$.
Moreover, we use $\delta_E(A)$ or $\delta(A)$ to denote $E(A,V\sm{A})$;
$\delta(A)$ is called a \emph{cut} of $G$.
By a $p$-cut we mean a cut of size $p$.
We use
$G[A]$ to denote the subgraph of $G$ induced by $A$,
$G-A$  to denote the subgraph of $G$ induced by $V\sm{A}$, and
$G-F$ to denote the graph $(V,\; E\sm{F})$.
We may use relaxed notation for singleton sets, e.g.,
we may use $G-v$ instead of $G-\{v\}$, etc.
A multi-graph $G$ is called $k$-edge connected if $|V(G)|\ge2$ and for
every $F\subseteq E(G)$ of size $<k$, $G-F$ is connected.

\begin{fact} \label{prop:containment}
Let $A, B \subseteq V$ be a pair of crossing sets.
For any edge-set $F \subseteq \binom{V}{2}$ and any
$S \in \{A \cap  B, A \cup B, A \setminus B, B \setminus A\}$,
we have $\delta_F(S) \subseteq \delta_F(A) \cup \delta_F(B)$.
\end{fact}
\begin{proof}
By examining cases, we can show that
$e\in \delta_F(S) \:\implies\: e\in\delta_F(A) \textup{~or~} e\in\delta_F(B)$.
\end{proof}

For any function $f : 2^V \to \{0,1\}$ and any edge-set $F \subseteq E$,
we say that $S \subseteq V$ is \emph{violated w.r.t.\ $f$, $F$}
if $|\delta_F(S)| < f(S)$, i.e.,
if $f(S)=1$ and there are no $F$-edges in the cut $\delta(S)$.
We drop $f$ and $F$ when they are clear from the context. The next observation states that the violated sets w.r.t.\ any
\genun function $f$ and any ``augmenting'' edge-set $F$ form a
\genun family.

\begin{fact} \label{prop:fmoduloF}
Let $f : 2^V \to \{0,1\}$ be a \genun function and $F \subseteq E$
be an edge-set. Define the function $f' : 2^V \to \{0,1\}$ such
that $f'(S) = 1$ if and only if both $f(S) = 1$ and $\delta_F(S) =
\emptyset$ hold. Then, $f'$ is also \genunnogap.
\end{fact}
\begin{proof}
Consider $A,B\subsetneq{V}$ such that $f'(A)=1=f'(B)$.
Clearly, $f(A)=1=f(B)$. Moreover, for any
$S \in \{A \cap  B, A \cup B, A \setminus B, B \setminus A\}$,
we have $\delta_F(S) = \emptyset$, by Fact~\ref{prop:containment}.
Since $f$ is \genunnogap, there are at least two distinct sets
$S_1,S_2 \in \{A \cap  B, A \cup B, A \setminus B, B \setminus A\}$
with $f$-value one.
Then, we have $f'(S_1)=1=f'(S_2)$ (since $\delta_F(S_1)=\emptyset=\delta_F(S_2)$).
Hence, $f'$ is \genunnogap.
\end{proof}

{
Informally speaking, the next lemma states that \genun families are closed under complementation.

\newcommand{\bS}{\overline{S}}

\begin{lemma}\label{lem:genuncross}
Suppose that $\F\subseteq 2^V$ is a \genun family.
Then
	$\F^* := \F \; \bigcup \;\; \{(V\sm{A}) \,|\, A\in\F\}$
is a \genun family.
\end{lemma}

\begin{proof}
{
Consider any two sets $S_1,S_2\in\F^*$.
We map these two sets to two sets $A,B\in\F$,
by taking complements if needed.
Since $\F$ is a \genun family, at least two of the four sets 
$A\cup{B}, A\cap{B}, A\sm{B}, B\sm{A}$ belong to $\F$.
Moreover, we have $S\in\F\:\implies\:\bS\in\F^*$.
By examining a few cases, we can conclude that
at least two of the four sets 
$S_1\cup{S_2}, S_1\cap{S_2}, S_1\sm{S_2}, S_2\sm{S_1}$ belong to $\F^*$,
thus, showing that $\F^*$ is a \genun family.

Next, we present the case~analysis.
Consider any two sets $S_1,S_2\in\F^*$.

\begin{description}
\item[\textbf{(1)}] $S_1\in\F,\;S_2\in\F$~:
\\
Since $\F$ is a \genun family, at least two of the four sets
$S_1\cup{S_2}, S_1\cap{S_2}, S_1\sm{S_2}, S_2\sm{S_1}$ belong to $\F^*\supseteq\F$.

\medskip

\item[\textbf{(2)}] $\bS_1\in\F,\;\bS_2\in\F$~:
\\
Let $A=\bS_1$ and let $B=\bS_2$.
Since $\F$ is a \genun family, at least two of the four sets
	$A\cup{B}, A\cap{B}, A\sm{B}, B\sm{A}$ belong to $\F$.
Observe the following:
\begin{itemize}
\item[] $A\cap{B} = \bS_1 \cap \bS_2$; taking the complement, we have
	$\overline{A\cap{B}} = S_1\cup{S_2}$; 
	hence, $A\cap{B}\in\F\:\implies\:{S_1}\cup{S_2}\in\F^*$.
\item[] $A\cup{B} = \bS_1 \cup \bS_2$; taking the complement, we have
	$\overline{A\cup{B}} = S_1\cap{S_2}$; 
	hence, $A\cup{B}\in\F\:\implies\:{S_1}\cap{S_2}\in\F^*$.
\item[] $A\sm{B} = \bS_1 \sm \bS_2 = S_2 \sm S_1$;
	hence, $A\sm{B}\in\F\:\implies\:{S_2}\sm{S_1}\in\F^*$.
\item[] $B\sm{A} = \bS_2 \sm \bS_1 = S_1 \sm S_2$;
	hence, $B\sm{A}\in\F\:\implies\:{S_1}\sm{S_2}\in\F^*$.
\end{itemize}
Hence, at least two of the four sets
$S_1\cup{S_2}, S_1\cap{S_2}, S_1\sm{S_2}, S_2\sm{S_1}$ belong to $\F^*$.

\medskip

\item[\textbf{(3)}] $\bS_1\in\F,\;S_2\in\F$~:
\\
Let $A=\bS_1$ and let $B=S_2$.
Since $\F$ is a \genun family, at least two of the four sets
	$A\cup{B}, A\cap{B}, A\sm{B}, B\sm{A}$ belong to $\F$.
Observe the following:
\begin{itemize}
\item[] $A\cap{B} = \bS_1 \cap S_2 = S_2 \sm S_1$
	hence, $A\cap{B}\in\F\:\implies\:{S_2}\sm{S_1}\in\F^*$.
\item[] $A\cup{B} = \bS_1 \cup S_2 = \overline{S_1 \sm S_2}$; taking the complement, we have
	$\overline{A\cup{B}} = S_1\sm{S_2}$; 
	hence, $A\cup{B}\in\F\:\implies\:{S_1}\sm{S_2}\in\F^*$.
\item[] $A\sm{B} = \bS_1 \sm S_2 = \overline{S_1 \cup S_2}$; taking the complement, we have
	$\overline{A\sm{B}} = S_1\cup{S_2}$; 
	hence, $A\sm{B}\in\F\:\implies\:{S_1}\cup{S_2}\in\F^*$.
\item[] $B\sm{A} = S_2 \sm \bS_1 = S_1 \cap S_2$;
	hence, $B\sm{A}\in\F\:\implies\:{S_1}\cap{S_2}\in\F^*$.
\end{itemize}
Hence, at least two of the four sets
$S_1\cup{S_2}, S_1\cap{S_2}, S_1\sm{S_2}, S_2\sm{S_1}$ belong to $\F^*$.
\end{description}
}
\end{proof}
}

\subsection*{Reducing the hitting set problem to the {$f$-connectivity~problem}}

{The following reduction from the hitting set problem
(which has a logarithmic hardness-of-approximation threshold \cite{WS-book}) to the
{$f$-connectivity~problem} shows that the (unrestricted)
{$f$-connectivity~problem} has a logarithmic hardness-of-approximation threshold.}

In the hitting~set problem, we are given a ground-set $X$ and a
family $\Ss$ of subsets of $X$;
the goal is to find a smallest subset $W$ of $X$ such that $W\cap{A}$
is non-empty for each set $A\in\Ss$; in other words, the goal is
to find a minimum-size cover of the sets of $\Ss$.

The hitting~set problem reduces to an $\fcproblem$ on a bipartite
graph $G=(L\cup{R},E)$ where we define $L := \{\ell_x : x \in
X\}$, $R := \{r_x : x \in X\}$, and $E := \{e_x = \ell_x r_x : x
\in X\}$; observe that $E$ is a perfect matching of $G$.
We define $f$ to be the indicator function of the
family $\{\{\ell_x : x \in A\} : A \in \Ss\}$.

Thus, each element $x\in{X}$ is modeled by an (isolated) edge
$e_x = \ell_x r_x$ of the bipartite graph,
and each set $A\in\Ss$ is modeled by fixing $f(\{\ell_x : x \in A\})$ to be~one.
Clearly, any solution $W\subseteq{X}$ of the hitting~set problem
corresponds to a solution $\{\ell_x r_x : x\in{W}\}$ of the $\fcproblem$ of the same size.

\subsection{The WGMV Primal-Dual Algorithm for Uncrossable Functions} \label{sec:wgmv}
{
In this section, we give a brief description of the primal-dual
algorithm of Williamson et al.~\cite{WGMV95} that achieves approximation
ratio~2 for any $\fcproblem$ such that $f$ is an uncrossable function.

\begin{theorem}[Lemma~2.1 in \cite{WGMV95}] \label{thm:primal-dual-wgmv}
Let $f : 2^V \to \{0,1\}$ be an uncrossable function. Suppose we
have a subroutine that for any given $F \subseteq E$, computes all
minimal violated sets w.r.t.\ $f$, $F$.
Then, in polynomial time and using a polynomial number of calls to
the  subroutine, we can compute a $2$-approximate solution to the given
instance of the $f$-connectivity problem.
\end{theorem}

The algorithm and its analysis are based on the following LP
relaxation of \eqref{eq:fconnectivityIP} (stated on the left) and its dual.
Define $\mcS := \{ S \subseteq V : f(S) = 1\}$.

\medskip

\noindent
\begin{minipage}[t]{0.50\textwidth} \label{eq:primaldualLPs}
\begin{center}
    \textbf{Primal LP}
\end{center}
\begin{align}
    & \qquad \, \, \, \min \quad \sum_{e \in E} c_e x_e & \notag \\
    & \text{subject to:} \sum_{e \in \delta(S)} x_e \geq 1 \quad \forall S \in \mcS \notag \\
    & \qquad \qquad \quad  0 \leq x_e \leq 1 \quad \, \, \, \forall e \in E \notag
\end{align}
\end{minipage}
\quad \vline \quad 
\begin{minipage}[t]{0.45\textwidth}
\begin{center}
    \textbf{Dual LP}
\end{center}
\begin{align}
& \qquad \, \, \max \quad \, \, \, \sum_{S \in \mcS} y_S \notag \\
& \text{subject to:} \sum_{S \in \mcS : e \in \delta(S)} y_S \leq c_e \quad \forall e \in E \notag \\
& \qquad \qquad \qquad y_S \geq 0 \qquad \qquad \, \, \, \forall S \in \mcS \notag
\end{align}
\end{minipage}

\medskip

The algorithm starts with an infeasible primal solution $F =
\emptyset$, which corresponds to $x = \chi^F = \mathbf{0} \in
\{0,1\}^E$, and a feasible dual solution $y = \mathbf{0}$.
At any time, we say that $S \in \mcS$ is \emph{violated} if
$\delta_F(S) = \emptyset$, i.e., the primal-covering constraint for
$S$ is not satisfied.
We call inclusion-wise minimal violated sets as \emph{active sets}.
An edge $e \in E$ is said to be \emph{tight} if $\sum_{S \in \mcS
: e \in \delta(S)} y_S = c_e$, i.e., the dual-packing constraint
for $e$ is tight.
Throughout the algorithm, the following conditions are maintained:
(i)~integrality of the primal solution; (ii)~feasibility of the
dual solution; (iii)~$y_S$ is never decreased for any $S$; and
(iv)~$y_S$ may only be increased for $S \in \mcS$ that are active.

The algorithm has two stages. In the first stage, the algorithm
iteratively improves primal feasibility by
{including tight edges in $F$ that are
	contained in the cut $\delta(S)$ of any active set $S$}.
If no such edge exists, then the algorithm uniformly increases $y_S$
for all active sets $S$ until a new edge becomes tight.
The first stage ends when $x = \chi^F$ becomes feasible.
In the second stage, called \emph{reverse delete}, the algorithm
removes redundant edges from $F$. Initially $F' = F$. The algorithm
examines edges picked in the first stage in reverse order, and
discards edges from $F'$ as long as feasibility is maintained. Note
that $F'$ is feasible if the subroutine in the hypothesis of
Theorem~\ref{thm:primal-dual-wgmv} does not find any (minimal)
violated sets.

The analysis of the $2$-approximation ratio is based on showing
that a relaxed form of the complementary slackness conditions hold on
``average''. Let $F'$ and $y$ be the primal and dual solutions
returned by the algorithm.
By the design of the algorithm, $\sum_{S \in \mcS: e \in  \delta(S)}
y_S = c_e$ holds for any edge $e \in F'$. Thus, the cost of $F'$
can be written as
$\sum_{e \in F'} \sum_{S \in \mcS : e \in \delta(S)} y_S =
	\sum_{S \in \mcS} y_S \cdot |\delta_{F'}(S)|$.
Observe that the approximation ratio follows from showing that
the algorithm always maintains the following inequality:
\begin{equation} \label{eq:primaldualanalysis} 
\sum_{S \in \mcS} y_S \cdot |\delta_{F'}(S)| \leq 2 \sum_{S \in \mcS} y_S.
\end{equation}

Consider any iteration and recall that the dual variables corresponding
to active sets were uniformly increased by an $\varepsilon > 0$
amount, until some edge became tight.
Let $\C$ denote the collection of active sets during this iteration.
During this iteration, the left-hand side of
\eqref{eq:primaldualanalysis} increases by $\varepsilon \cdot \sum_{S
\in \C} |\delta_{F'}(S)|$ and the right-hand side increases by $2
\cdot \varepsilon \cdot |\C|$. Thus, \eqref{eq:primaldualanalysis}
is maintained if one can show that the average $F'$-degree of active
sets in {every} iteration is $\leq2$, and this forms the crux of
the WGMV analysis.

We refer the reader to \cite{GW-bookchapter} for a detailed discussion
of the primal-dual method for network design problems.

}
}
{
\section{Extending the WGMV Primal-Dual Method to \GenUn functions
	{with property~\propred}} \label{sec:general-WGMV}
In this section, we prove our main result, Theorem~\ref{thm:pliablemain}:
we show that the primal-dual algorithm outlined in Section~\ref{sec:wgmv}
is a $\pliableapx$-approximation algorithm for the $\fcproblem$
where $f$ is a \genun function with property~\propred.
Our analysis follows the same high-level plan as that of
Williamson et al.~\cite{WGMV95} which was outlined in
Section~\ref{sec:wgmv}. We will show that, in any iteration of the
first stage of the primal-dual algorithm,
{\[ \sum_{C\in \C} |\delta_{F'}(C)| \leq \pliableapx |\C|,\]}
where $\C$ is the collection of
active sets in that iteration, and $F'$ is the set of edges output
by the algorithm at termination, i.e., after the reverse delete stage.

For the remainder of this proof we assume that the iteration, and
thus $\C$, is fixed. We define $H := \cup_{C\in\C} \delta_{F'}(C)$.
(Informally speaking, $H$ is the subset of $F'$ that is relevant
for the analysis of our fixed iteration.)
Additionally,
{for notational convenience},
we say that two sets $A,B$ \emph{overlap} if $A\setminus
B,A\cap B$ and $B\setminus A$ are all non-empty.
(Clearly, if $A,B$ cross, then $A,B$ overlap;
if $A\cup{B}=V$, then $A,B$ do not cross but $A,B$ could overlap.)

\medskip

{
We begin with a lemma that can be proved by the same arguments as in the proof of
\cite[Lemma~5.1]{WGMV95}.

\begin{lemma}\label{lem:witnessfamily}
For any edge $e\in H:= \cup_{C\in\C} \delta_{F'}(C)$, there exists a witness set $S_e \subseteq V$ such that: 
\begin{enumerate}[(i)]
    \item $f(S_e) = 1$ and $S_e$ is violated in the current iteration, and
    \item $\delta_{F'}(S_e) = \{e\}$.
\end{enumerate}
\end{lemma}
}


\medskip

Our proof of the following key lemma is presented in Appendix~\ref{append:proofsec3}.

\begin{lemma}
\label{lem:lamfamily}
{
There exists a laminar family of witness sets.
}
\end{lemma}

{
\begin{lemma}
The active sets in $\C$ are pair-wise disjoint.
\end{lemma}
\begin{proof}
{Consider two sets $C_1,C_2 \in \C$ such that $C_1\cap{C_2}\not=\emptyset$.}
{Then by the definition of \genun functions},
one of the sets
$C_1\cap C_2$, $C_1\sm{C_2}$, or $C_2\sm{C_1}$ is violated; thus,
a proper subset of either $C_1$ or $C_2$ is violated.  This is a
contradiction because $C_1$ and $C_2$ are minimal violated sets and
no proper subset of $C_1$ (respectively, $C_2$) is violated.
\end{proof}

Let $\Ls$ be the laminar family of witness sets together with the node-set $V$.
Let $\T$ be a rooted tree that represents $\Ls$;
for each set $S\in\Ls$, there is a node $v_S$ in $\T$,
and the node $v_V$ is taken to be the root of $\T$.
{Thus, $\T$ has an edge $v_Qv_S$ iff $Q$ is the smallest set of $\Ls$
that properly contains the set $S$ of $\Ls$.}
Let $\psi$ be a mapping from $\C$ to $\Ls$ that
maps each active set $C$ to the smallest set $S\in\Ls$ that contains it.
If a node $v_S$ of $\T$ has some active set mapped to its associated set $S$, then we
call $v_S$ \emph{active} and we assign the color red to $v_S$.
Moreover, we assign the color green to each of the non-active nodes
of $\T$ that are incident to three or more edges of $\T$;
thus, node $v_S$ of $\T$ is green iff $\deg_{\T}(v_S)\geq3$ and $v_S$ is not active.
Finally, we assign the color black to each of the remaining nodes of $\T$;
thus, node $v_S$ of $\T$ is black iff $\deg_{\T}(v_S)\leq2$ and $v_S$ is not active.

Let the number of red, green and black nodes of $\T$ be denoted by
$n_R, n_G$ and $n_B$, respectively. Clearly,
{$n_R+n_G+n_B = |\T| = |H|+1$}.
Let $n_L$ denote the number of leaf nodes of $\T$.

\begin{lemma} \label{lem:tree-count}
The following statements hold:
\begin{enumerate}[(i)]
\item Each leaf node of $\T$ is red.
\item We have $n_G \leq n_L \leq n_R { \leq |\C|} $.
\end{enumerate}
\end{lemma}
\begin{proof}
The first statement follows by repeating the argument in \cite[Lemma~5.3]{WGMV95}.

{We have $n_G \leq n_L$ because the number of leaves in
any tree is at least the number of nodes that are incident to three
or more edges of the tree.
Moreover, by~(i), we have $n_L \leq n_R$. 
Every red node of $\T$ is associated with a set in $\C$ by the mapping $\psi$,
hence, $n_R \leq |\C|$.}
\end{proof}

Observe that each black node of $\T$ is incident to two edges of $\T$;
thus, every black non-root node of $\T$ has a unique child.

Let us sketch our plan for proving Theorem~\ref{thm:pliablemain}.
Clearly, the theorem would follow from the inequality
$\sum_{C\in\C}|\delta_{F'}(C)| \;\leq\; \pliableapx \; |\C|$;
thus, we need to prove an upper~bound of $\pliableapx \; |\C|$ on the number
of ``incidences'' between the edges of $F'$ and the cuts $\delta(C)$
of the active sets $C\in\C$.
We start by assigning a token to $\T$ corresponding to each
``incidence''. In more detail, for each edge $e\in F'$ and cut
$\delta(C)$ such that $C\in\C$ and $e\in\delta(C)$ we assign one
token to the node $v_{S_e}$ of $\T$ that represents the witness set
$S_e$ of the edge $e$.
Thus, the total number of tokens assigned to $\T$ is
$\sum_{C\in\C}|\delta_{F'}(C)|$; moreover, after the initial
assignment, it can be seen that each node of $\T$ has $\leq2$ tokens
(see Lemma~\ref{lem:twotokens} below).
Then we redistribute the tokens according to a simple rule such that
(after redistributing) each of the red/green nodes has $\leq8$ tokens
and each of the black nodes has no tokens.
Lemma~\ref{lem:colorednodes} (below) proves this key claim by applying property~\propred.
The key claim implies that the total number of tokens assigned to $\T$ is
$\leq 8n_R + 8n_G \leq 16n_R \leq 16|\C|$ (by Lemma~\ref{lem:tree-count}).
This concludes our sketch.

We apply the following two-phase scheme to assign tokens to the nodes of $\T$.
\begin{itemize}
\item In the first phase, for $C\in \C$ and $e\in \delta_{F'}(C)$,
we assign a new token to the node $v_{S_e}$ corresponding to the
witness set $S_e$ for the edge $e$.
At the end of the first phase, observe that the root $v_V$ of $\T$
has no tokens (since the set $V$ cannot be a witness set).

\item In the second phase, we apply a root-to-leaves scan of $\T$ (starting from the root $v_V$).
Whenever we scan a black node, then we move all the tokens at that node
to its unique child node.
(There are no changes to the token distribution when we scan a red node or a green node.)
\end{itemize}

\begin{lemma}\label{lem:twotokens}
At the end of the first phase, each node of $\T$ has $\leq2$ tokens.
\end{lemma}
\begin{proof}
Consider a non-root node $v_{S_e}$ of $\T$.
This node corresponds to a witness set $S_e\in\Ls$ and
$e$ is the unique edge of $F'$ in $\delta(S_e)$.
The edge $e$ is in $\leq2$ of the cuts $\delta(C), C\in\C,$
because the active sets are pairwise disjoint
(in other words, the number of ``incidences'' for $e$ is $\leq2$).
No other edge of $F'$ can assign tokens to $v_{S_e}$ during the first phase.
\end{proof}

\begin{lemma}\label{lem:colorednodes}
The following statements hold:
\begin{enumerate}[(i)]
\item
{Consider any path of $\T\sm\{v_V\}$ with four nodes
$v_{S_4}, v_{S_3}, v_{S_2}, v_{S_1}$
such that $S_1 \subsetneq S_2 \subsetneq S_3 \subsetneq S_4$.
At least one of the four nodes is red or green (i.e., non-black).
}

\item Hence, after token redistribution, each red or green node of $\T$ has $\leq 8$ tokens
and each black node of $\T$ has zero tokens.
\end{enumerate}
\end{lemma}
\begin{proof}
{
{For the sake of contradiction, assume that there exists a
path of $\T\sm\{v_V\}$ with four black nodes
$v_{S_4}, v_{S_3}, v_{S_2}, v_{S_1}$
such that $S_1 \subsetneq S_2 \subsetneq S_3 \subsetneq S_4$;
clearly, $S_1,S_2,S_3,S_4$ are witness sets {in} $\Ls$.
}
For $i\in\{1,2,3,4\}$, let $S_i$ be the witness set of edge $e_i=\{a_i,b_i\}\in{F'}$;
note that $e_i$ has exactly one end-node in $S_i$, call it $a_i$.
Clearly, for $i\in\{1,2,3\}$, both nodes $a_i,b_i$ are in $S_{i+1}$ 
(since $e_{i+1}$ is the unique edge of $F'$ in $\delta(S_{i+1})$).

{
\begin{figure}[htb] \centering
{
\begin{tikzpicture} 
\begin{scope}[every node/.style={circle, fill=black, draw, inner sep=0pt, minimum size = 0.15cm, label distance=0.1cm}]
	\node[draw=none,fill=none]	(v04) at ($(0,0)+(-6,1.5)$) {};
	\node[draw=none,fill=none]	(v03) at ($(0,0)+(-5,1.5)$) {};
	\node[draw=none,fill=none]	(v02) at ($(0,0)+(-4,1.5)$) {};
	\node[draw=none,fill=none]	(v01) at ($(0,0)+(-3,1.5)$) {};
	\node[draw=none,fill=none]			(v1) at ($(0,0)+(1,1.5)$) {};
	\node[draw=none,fill=none]			(v2) at ($(0,0)+(2,1.5)$) {};
	\node[draw=none,fill=none]			(v3) at ($(0,0)+(3,1.5)$) {};
	\node[draw=none,fill=none]			(v4) at ($(0,0)+(4,1.5)$) {};
	\node[draw=none,fill=none,label={[label distance=0.0]0:$S_1$}] (s1) at ($(0,0)+(-2.25,1.5)$) {};
	\node[label={[label distance=0.0]180:$a_1$}]	(a1) at ($(0,0)+(-0.5,1.5)$) {};
	\node[label={[label distance=0.0]270:$b_1$}]	(b1) at ($(0,0)+(+0.5,1.5)$) {};
	\node[draw=none,fill=none,label={[label distance=0.0]0:$S_2$}] (s2) at ($(0,0)+(-3.00,1.5)$) {};
	\node[label={[label distance=0.0]135:$a_2$}]	(a2) at ($(0,0)+(+0.5,1.75)$) {};
	\node[label={[label distance=0.0]270:$b_2$}]	(b2) at ($(0,0)+(+1.5,1.75)$) {};
	\node[draw=none,fill=none,label={[label distance=0.0]0:$S_3$}] (s3) at ($(0,0)+(-4.00,1.5)$) {};
	\node[label={[label distance=0.0]135:$a_3$}]	(a3) at ($(0,0)+(+1.5,2.00)$) {};
	\node[label={[label distance=0.0]270:$b_3$}]	(b3) at ($(0,0)+(+2.5,2.00)$) {};
	\node[draw=none,fill=none,label={[label distance=0.0]0:$S_4$}] (s4) at ($(0,0)+(-5.00,1.5)$) {};
	\node[label={[label distance=0.0]180:$a_4$}]	(a4) at ($(0,0)+(+2.0,2.50)$) {};
	\node[label={[label distance=0.0]90:$b_4$}]	(b4) at ($(0,0)+(+3.0,2.50)$) {};
\end{scope}

\begin{scope}[line width=0.6pt]
	\draw[thick,dashed] ($0.5*(v01)+0.5*(v1)$) ellipse (1.25cm and 0.60cm);

	\draw[thick,dashed] ($0.5*(v02)+0.5*(v2)$) ellipse (2.00cm and 1.00cm);

	\draw[thick,dashed] ($0.5*(v03)+0.5*(v3)$) ellipse (3.00cm and 1.50cm);

	\draw[thick,dashed] ($0.5*(v04)+0.5*(v4)$) ellipse (4.00cm and 2.00cm);
\end{scope}

        \begin{scope}[every edge/.style={draw=black}]
                                
                \path[thick] (a1) edge[] node {} (b1);  
                \path[thick] (a2) edge[] node {} (b2);  
                \path[thick] (a3) edge[] node {} (b3);  
                \path[thick] (a4) edge[] node {} (b4);  
        \end{scope}
\end{tikzpicture}
}
\caption{
        \label{f:lem-colorednodes}
	Illustration of the witness~sets
	$S_1 \subsetneq S_2 \subsetneq S_3 \subsetneq S_4$ and the
	edges $a_ib_i\; (i=1,\dots,4)$, in the proof of Lemma~\ref{lem:colorednodes}.
}               
\end{figure}
}               

Let $C\in\C$ be an active set such that $e_1\in\delta(C)$.

\begin{claim} \label{clm:colorednodes.1}
$C$ is not a subset of $S_1$.
\end{claim}

For the sake of contradiction, suppose that $C$ is a subset of $S_1$.
Since $e_1$ has (exactly) one end-node in $C$ and $b_1\not\in S_1$,
we have $a_1\in{C}$.
Let $W$ be the smallest set in $\Ls$ that contains $C$.
Then $W\subseteq S_1$, and, possibly, $W=S_1$.
Thus, we have $a_1\in W$ and $b_1\not\in W$, hence, $e_1\in\delta(W)$.
Then we must have $W=S_1$
(since $e_1$ is in exactly one of the cuts $\delta(S), S\in\Ls$).
Then the mapping $\psi$ from $\C$ to $\Ls$ maps $C$ to $W=S_1$,
hence, $v_{S_1}$ is colored red. This is a contradiction.

\begin{claim} \label{clm:colorednodes.2}
$C$ crosses each of the sets $S_2, S_3, S_4$.
\end{claim}

First, observe that $e_1$ has (exactly) one end-node in $C$ and has
both end-nodes in $S_2$.
Hence, both $S_2\cap{C}$ and $S_2\sm{C}$ are non-empty.
Next, using Claim~\ref{clm:colorednodes.1}, we can prove that $C$ is not a subset of $S_2$.
$\big($Otherwise, $S_2$ would be the smallest set in $\Ls$
that contains $C$, hence, $v_{S_2}$ would be colored red.$\big)$
Repeating the same argument, we can prove that $C$ is not a subset of $S_3$,
and, moreover, $C$ is not a subset of $S_4$.
Finally, note that $V\sm(C\cup{S_4})$ is non-empty.
$\big($Otherwise,
{at least one of $C\setminus S_4$ or $C\cap S_4$ would be violated, since $f$ is a \genunnogap\ function},
and that would contradict the fact that $C$ is a minimal violated set.$\big)$
Observe that $S_2$ crosses $C$ because all four sets
$S_2\cap{C}$, $S_2\sm{C}$, $C\sm{S_2}$, $V\sm(S_2\cup{C})$ are non-empty
(in more detail, we have
$|\{a_1,b_1\} \cap (S_2\cap{C})|=1$, $|\{a_1,b_1\}\cap (S_2\sm{C})|=1$, 
$C\not\subseteq{S_2}\implies C\sm{S_2}\not=\emptyset$,
$V\sm(C\cup{S_2})\supseteq V\sm(C\cup{S_4})\not=\emptyset$).
Similarly, it can be seen that $S_3$ crosses $C$, and $S_4$ crosses $C$.

\begin{claim} \label{clm:colorednodes.3}
Either $S_3\sm(C\cup{S_2})$ is non-empty or
       $S_4\sm(C\cup{S_3})$ is non-empty.
\end{claim}

For the sake of contradiction, suppose that both sets
$S_3\sm(C\cup{S_2})$,
$S_4\sm(C\cup{S_3})$  are empty.
Then $C\supseteq S_4\sm{S_3}$ and $C\supseteq S_3\sm{S_2}$.
Consequently, both end-nodes of $e_3$ are in $C$
(since $a_3\in S_3\sm{S_2}$ and $b_3\in S_4\sm{S_3}$).
This leads to a contradiction, since $e_3\in F'$ is incident
to an active set in $\C$, call it $C_3$ (i.e., $e_3\in\delta(C_3)$),
hence, one of the end-nodes of $e_3$ is in both $C$ and $C_3$,
whereas the active sets are pairwise disjoint.

To conclude the proof of the lemma, suppose that 
$S_4\sm(C\cup{S_3})$  is non-empty (by Claim~\ref{clm:colorednodes.3});
the other case, namely, $S_3\sm(C\cup{S_2})\not=\emptyset$, can be
handled by the same arguments.
Then, by property~\propred, $S_4\sm(C\cup{S_3})$ is a violated set,
therefore, it contains a minimal violated set, call it $\wt{C}$.
Clearly, the mapping $\psi$ from $\C$ to $\Ls$ maps the active set $\wt{C}$
to a
{witness set $S_{\wt{C}}$ which is the smallest set in $\Ls$ that contains $\wt{C}$}.
Either $S_{\wt{C}} = S_4$ or else $S_{\wt{C}}$ is a subset of of $S_4\sm{S_3}$.
Both cases give contradictions;
in the first case, $S_4$ is colored red,
and in the second case, $v_{S_4}$ has $\geq2$ children in $\T$ so that
$S_4$ is colored either green or red.
Thus, we have proved the first part of the lemma.

The second part of the lemma follows by Lemma~\ref{lem:tree-count} and
the sketch given below Lemma~\ref{lem:tree-count}.
In more detail, at the start of the second phase, each node of $\T$
has $\leq2$ tokens, by Lemma~\ref{lem:twotokens}.
In the second phase, we redistribute the tokens such that each
(non-root) black node ends up with no tokens, and each red/green
node $v_S$ receives $\leq 6$ redistributed tokens because there are
$\leq3$ black ancestor nodes of $v_S$ that could send their tokens
to $v_S$ (by the first part of the lemma).
Hence, each non-root non-black node has $\leq8$ tokens, after token redistribution. 
}
\end{proof}
}
{Theorem~\ref{thm:pliablemain} follows from
	Lemmas~\ref{lem:witnessfamily}--\ref{lem:colorednodes}.}
}
{
\section{\boldmath {$O(1)$}-Approximation Algorithm for Augmenting Small Cuts} \label{sec:nearmincuts}
In this section, we give a ${\pliableapx}$-approximation algorithm for the
$\augsmallcuts$ problem, thereby proving
Theorem~\ref{thm:nearmincutsaugmentation}. Our algorithm for
$\augsmallcuts$ is based on a reduction to an instance of the
$f$-connectivity problem on the graph $H = (V,L)$ for a \genun 
function $f$ with property~\propred.

Recall the $\augsmallcuts$ problem:
we are given an undirected graph $G = (V,E)$ with edge-capacities
$u \in \Qp^E$, a set of links $L \subseteq \binom{V}{2}$ with costs
$c \in \Qp^L$, and a threshold $\capbound \in \Qp$.  A subset $F
\subseteq L$ of links is said to \emph{augment} a node-set $S$ if
there exists a link $e \in F$ with exactly one end-node in $S$.
The objective is to find a minimum-cost $F\subseteq{L}$ that augments all
non-empty $S \subsetneq V$ with $u(\delta_E(S)) < \capbound$.

\begin{proofof}{Theorem~\ref{thm:nearmincutsaugmentation}}
Define $f : 2^V \to \{0,1\}$ such that $f(S) = 1$ if and only if
$S \notin \{\emptyset,V\}$ and $u(\delta_E(S)) < \capbound$. We
apply Theorem~\ref{thm:pliablemain} for the $f$-connectivity problem
on the graph $H = (V,L)$ with edge-costs $c \in \Qp^L$ to obtain
a ${\pliableapx}$-approximate solution $F \subseteq L$. By our choice of
$f$, there is a one-to-one cost-preserving correspondence between
feasible augmentations for $\augsmallcuts$ and feasible solutions
to the $f$-connectivity problem. Thus, it remains to argue that the
assumptions of Theorem~\ref{thm:pliablemain} hold.

First, we show that $f$ is \genunnogap. Note that $f$ is symmetric
and $f(V) = 0$. Consider sets $A, B \subseteq V$ with $f(A)
= f(B) = 1$.
By submodularity and symmetry of cuts in undirected graphs,
we have: $\max\{u(\delta(A \cup B)) + u(\delta(A
\cap B)),\; u(\delta(A\sm{B})) + u(\delta(B\sm{A}))\}
\leq u(\delta(A)) + u(\delta(B))$. Since the right hand side is
strictly less than $2 \capbound$, we have $f(A \cap B) + f(A \cup
B) \geq 1$ and $f(A \setminus B) + f(B \setminus A) \geq 1$,
hence, $f$ is \genunnogap.

Second, we argue that $f$ satisfies property~\propred.
Fix some edge-set $F \subseteq L$, and define $f' : 2^V \to \{0,1\}$
such that $f'(S) = 1$ if and only if $f(S) = 1$ and $\delta_F(S) =
\emptyset$. By Fact~\ref{prop:fmoduloF}, $f'$ is also
\genunnogap.  Consider sets $C, S_1, S_2 \subseteq V$, $S_1\subsetneq{S_2}$,
that are violated w.r.t.\ $f$, $F$, i.e., $f'(C) = f'(S_1) = f'(S_2) = 1$.
Further, suppose that $C$ is minimally violated, and $C$ crosses
both $S_1$ and $S_2$.
Suppose that $S_2 \setminus (S_1 \cup C)$ is non-empty (the other case is trivial).
To show that 
$S_2 \setminus (S_1 \cup C)$
is violated w.r.t.\ $f,\;F$, we have to show that
(i)~$\delta_F(S_2 \setminus (S_1 \cup C))$ is empty
and
(ii)~$u(\delta_E(S_2 \setminus (S_1 \cup C))) < \capbound$.
Observe that $S_2$ crosses $(S_1\cup{C})$.
To show~(i), we apply Fact~\ref{prop:containment} twice;
first, we show that $\delta_F(S_1\cup{C})$ is empty (since $\delta_F(C),\, \delta_F(S_1)$ are empty),
and then we show that $\delta_F(S_2 \setminus (S_1 \cup C))$ is empty (since $\delta_F(S_2)$ is empty).
{To show~(ii), observe that
{the multiset~}
	$\delta_E(S_2 \setminus (S_1\cup C)) \, \cup \, \delta_E(C \setminus S_2)$
{~is a subset of the multiset~}
	$\delta_E(S_2) \, \cup \, \delta_E(C\cup S_1)$,
{~and,}
{for each edge, its multiplicity in~}
	$\delta_E(S_2 \setminus (S_1\cup C)) \, \cup \, \delta_E(C \setminus S_2)$
{~is $\leq$ its multiplicity in~}
	$\delta_E(S_2) \, \cup \, \delta_E(C\cup S_1)$.
}
(Note that for disjoint sets $A_1,A_2,A_3\subsetneq{V}$,
{the multiset} $\delta(A_1) \cup \delta(A_2)$ is a subset of
{the multiset} $\delta(A_1\cup{A_3}) \cup \delta(A_2\cup{A_3})$.)
Moreover, we claim that $u(\delta_E(C\cup S_1)) < \capbound$
and $u(\delta_E(C \setminus S_2)) \geq \capbound$.
The two claims immediately imply~(ii) (since $u(\delta_E(S_2)) < \capbound$).

Next, we prove the two claims.
Note that the sets $C\cap{S_1}, C\sm{S_1}, S_1\sm{C}, V\sm(C\cup{S_1})$
are non-empty, and note that $f'(C\cap{S_1})=0=f'(C\sm{S_1})$ since
$C$ is a minimal violated set.
Since $f'$ is \genunnogap\ and $f'(C)=1=f'(S_1)$, we have $f'(C\cup{S_1})=1$. 
By Fact~\ref{prop:containment}, $\delta_F(C \cup S_1) = \emptyset$, hence, $f(C \cup S_1) = 1$;
equivalently, $u(\delta_E(C \cup S_1)) < \capbound$.
Since $C$ is a minimal violated set, $f'(C\sm{S_2})=0$.
Moreover, $\delta_F(C\sm{S_2}) = \emptyset$, by Fact~\ref{prop:containment}. 
Hence, $f(C \sm S_2) = 0$;
equivalently, $u(\delta_E(C \sm S_2)) \geq \capbound$.

Last, we describe a polynomial-time subroutine that for any $F
\subseteq L$ gives the collection of all minimal violated sets w.r.t.\ $f$, $F$.
Assign a capacity of $\capbound$ to all edges in
$F$, and consider the graph $G' = (V,E')$ where $E' := E \cup F$.
Then, the family of minimal violated sets is given by $\{ \emptyset
\subsetneq S \subsetneq V : u(\delta_{E'}(S)) < \capbound, \,
u(\delta_{E'}(A)) \geq \capbound \; \forall \; \emptyset \subsetneq
A \subsetneq S \}$.
We use the notion of solid sets to find all such minimally violated
sets; see Naor, Gusfield, and Martel \cite{NGM97} and see Frank's book \cite{Frank2011}.
A solid set of an undirected graph $H = (V,E'')$ with capacities
$w \in \Rp^{E''}$ on its edges is a non-empty node-set $Z \subsetneq
V$ such that $w(\delta_{E''}(X)) > w(\delta_{E''}(Z))$ for all
non-empty $X \subsetneq Z$.
{Note that the family of minimal violated sets of
interest to us is a sub-family of the family of solid sets of $G'$.}
The family of all solid sets of a graph can be listed in polynomial
time, see \cite{NGM97} and \cite[Chapter~7.3]{Frank2011}.  Hence, we
can find all minimal violated sets w.r.t.\ $f$, $F$ in polynomial time,
{by examining the list of solid sets to check (1)~whether there is
a solid set $S$ that is violated, and (2)~whether every proper subset
of $S$ that is a solid set is \emph{not} violated.}
This completes the proof of the theorem. \qedhere
\end{proofof}
}
{
\section{\boldmath {$O(k/u_{min})$}-Approximation Algorithm for the Capacitated {$k$}-Edge-Connected Subgraph Problem} \label{sec:capkecss}
In this section, we give a $\pliableapx \cdot {\lceil k/u_{\mathrm{min}} \rceil}$-approximation algorithm for the Cap-$k$-ECSS problem, thereby proving Theorem~\ref{thm:CapkECSS}.
Our algorithm is based on repeated applications of Theorem~\ref{thm:nearmincutsaugmentation}.

Recall the capacitated $k$-edge-connected subgraph problem (Cap-$k$-ECSS):
we are given an undirected graph $G = (V,E)$ with edge costs $c \in
\Qp^E$ and edge capacities $u \in \mathbb{Z}_{\geq 0}^E$. The goal is
to find a minimum-cost subset of the edges $F\subseteq E$ such that the
capacity across any cut in $(V,F)$ is at least $k$, i.e.,
$u(\delta_F(S)) \geq k$ for all non-empty sets $S\subsetneq V$.

\begin{proofof}{Theorem~\ref{thm:CapkECSS}}
The algorithm is as follows: Initialize $F\,:=\,\emptyset$.
While the minimum capacity of a cut $\delta(S),\, \emptyset\neq{S}\subsetneq{V},$
in $(V,F)$ is less than $k$, run the approximation algorithm from
Theorem~\ref{thm:nearmincutsaugmentation} with input $G=(V,F)$ and
$L = E\sm{F}$, to augment all cuts
$\delta(S),\, \emptyset\neq{S}\subsetneq{V},$ with $u(\delta(S))<k$
and obtain a valid augmentation $F' \subseteq L$.
Update $F$ by adding $F'$, that is, $F\,:=\,F\cup{F'}$.
On exiting the while loop, output the set of edges $F$.

At any step of the algorithm, let $\lambda$ denote the minimum
capacity of a cut in $(V,F)$, i.e.,
$\lambda := \min\{ u(\delta_F(S)) \;:\; \emptyset \subsetneq S \subsetneq V \}$.

The above algorithm outputs a feasible solution since, upon exiting
the while loop, $\lambda$ is at least $k$.
Let $F^*\subseteq E$ be an optimal solution to the Cap-$k$-ECSS instance.
Notice that $F^*\sm{F}$ is a feasible choice for $F'$
during any iteration of the while loop. Hence, by
Theorem~\ref{thm:nearmincutsaugmentation}, $c(F')\leq
\pliableapx\cdot{c(F^*)}$. We claim that the above algorithm requires
at most $\lceil\frac{k}{u_{min}}\rceil$ iterations of the while loop.
This holds because each iteration of the while loop (except possibly
the last iteration) raises $\lambda$ by at least $u_{min}$.
(At the start of the last iteration, $k-\lambda$ could be less than $u_{min}$,
and, at the end of the last iteration, $\lambda$ could be equal to $k$).
Hence, at the end of the algorithm, $c(F)\leq \pliableapx\cdot
{\lceil\frac{k}{u_{min}}\rceil} c(F^*)$. This completes the proof. \qedhere
\end{proofof}

Our new result (Theorem~\ref{thm:nearmincutsaugmentation})
is critical for the bound of ${\lceil\frac{k}{u_{min}}\rceil}$ on the number of iterations of this algorithm. Earlier methods only allowed augmentations of minimum~cuts, so such methods may require as many as $\Omega(k)$ iterations.
(In more detail, the earlier methods would augment the cuts of $(V,F)$ of capacity $\lambda$ but would not augment the cuts of capacity $\geq \lambda + 1$; thus, cuts of capacity $\lambda+1$ could survive the augmentation step.)
}
{
\section{\boldmath {$O(1)$}-Approximation Algorithm for {$\ptwofgc$}}  \label{sec:p2fgc}
In this section, we present a $\oddapx$-approximation algorithm for
$\ptwofgc$, by applying our results from Section~\ref{sec:general-WGMV}.

Recall (from Section~\ref{sec:intro}) that the algorithmic goal in
$\ptwofgc$ is to find a minimum-cost edge-set $F$ such
that for any pair of unsafe edges $e,f\in F\cap\unsafe$, the subgraph
$(V,F\setminus{\{e,f\}})$ is $p$-edge~connected.

{
Our algorithm works in two stages. First, we compute a feasible
edge-set $F_1$ for $\ponefgc$ on the same input graph, by applying
the 4-approximation algorithm of \cite{BCHI23}.  We then
augment the subgraph $(V,F_1)$ using additional edges.
Since $F_1$ is a feasible edge-set for $\ponefgc$, any cut
$\delta(S),\ \emptyset \subsetneq S \subsetneq V$, in the subgraph $(V,F_1)$
either (i)~has at least $p$ safe edges or (ii)~has at least $p+1$ edges
(see below for a detailed argument).
Thus the cuts that need to be augmented have exactly $p+1$ edges
and contain at least two~unsafe edges.
Augmenting all {such} cuts by at least one (safe
or unsafe) edge will ensure that we have a feasible solution to $\ptwofgc$.

The following example shows that when $p$ is odd, then the function $f$ in
the $\fconn$ problem  associated with $\ptwofgc$ may \emph{not} be an uncrossable function.
\begin{example} \label{eg:4nodes-odd-p}
We construct the graph $G$ by starting with a 4-cycle $v_1, v_2,
v_3, v_4, v_1$ and then replacing each edge of the 4-cycle by a
pair of parallel edges; thus, we have a 4-regular graph with 8~edges;
we designate the following four edges as unsafe (and the other four
edges are safe): both copies of edge $\{v_1,v_4\}$, one copy of edge
$\{v_1,v_2\}$, and one copy of edge $\{v_3,v_4\}$.  Clearly, $G$
is a feasible instance of $(3,1)$-$\fgc$.  On the other hand, $G$
is infeasible for $(3,2)$-$\fgc$, and
{the sets $\{v_1,v_2\}$ and $\{v_2,v_3\}$ are violated}.
Note that the indicator function $f$ associated with the
{violated node-sets} is not uncrossable
{(observe that the sets $\{v_2\}$ and $\{v_3\}$ are not violated)}.
Moreover, observe that the minimal violated set $C=\{v_2,v_3\}$
crosses the violated set $S=\{v_1,v_2\}$.
\end{example}
}

\begin{proofof}{Theorem~\ref{thm:ptwofgc}}
In the following, we use $F$ to denote the set of edges picked by
the algorithm at any step of the execution; we mention that our
correctness arguments are valid despite this ambiguous notation;
moreover, we use $\delta(S)$ rather than $\delta_F(S)$ to refer to
a cut of the subgraph $(V,F)$, where $\emptyset\not=S\subseteq{V}$.

Since $F$ is a feasible edge-set for $\genfgc{p}{1}$,
any cut $\delta(S)$ (where $\emptyset\not=S\subseteq{V}$) either
(i)~has at least $p$ safe edges or (ii)~has  at least $p+1$ edges.
Consider a node-set $S$ that violates the requirements of the
$\ptwofgc$~problem.  We have $\emptyset\not=S\subsetneq{V}$ and
there exist two unsafe edges $e,f\in\delta(S)$ such that
	{$|\delta(S) \setminus \{e,f\}| \leq p-1$}.
Since $F$ is feasible for $\genfgc{p}{1}$, we have
$|\delta(S)\setminus\{e\}|\geq{p}$ and $|\delta(S)\setminus\{f\}|
\geq{p}$.
	{Thus, $|\delta(S)|=p+1$.}
In other words, the node-sets $S$ that need to be augmented have
exactly $p+1$ edges in $\delta(S)$, at least two of which are unsafe
edges.
Let $f:2^V\to\{0,1\}$ be the indicator function of these violated sets.
Observe that $f$ is symmetric, that is, $f(S)=f(V\sm{S})$ for any $S\subseteq{V}$;
this additional property of $f$ is useful for our arguments.
We claim that $f$ is a \genun function that satisfies property~\propred, hence, we
obtain an $O(1)$-approximation algorithm for $\ptwofgc$, via the
primal-dual method and Theorem~\ref{thm:pliablemain}.

Our proof of the following lemma is presented in \cite[Section~5]{BCGI:arxiv}
as well as in Appendix~\ref{append:proofsec6}

	\begin{lemma}
\label{lem:ptwofgc.1}
{
$f$ is a \genun function that satisfies property~\propred. Moreover, for even $p$,  $f$ is an uncrossable function. 
}
	\end{lemma}

Lastly, we show that there is a polynomial-time subroutine for
computing the minimal violated sets.
Consider the graph $(V,F)$.  Note that size of a minimum cut of
$(V,F)$ is at least $p$ since $F$ is a feasible edge-set for $\genfgc{p}{1}$.
The violated sets are subsets $S\subseteq V$ such that $\delta(S)$
contains exactly $p+1$ edges, at least two of which are unsafe edges.
Clearly, all the violated sets are contained in the
family of sets $S$ such that $\delta(S)$ is a 2-approximate min-cut of $(V,F)$; in other words,
$\{ S \subsetneq V \;:\: p \leq |\delta(S)| \leq 2p \}$ contains all the violated sets.
It is well known that the family of 2-approximate min-cuts in a
graph can be listed in polynomial time, see \cite{KS96,NNI97}.
Hence, we can find all violated sets and all minimally violated sets in polynomial time.

Thus, we have a $\oddapx$-approximation algorithm for $\ptwofgc$
via the primal-dual algorithm of \cite{WGMV95} based on our results
in Section~\ref{sec:general-WGMV}.
In more detail,
we first find a solution to a $\ponefgc$ instance of cost $\leq{4}\opt$,
and then, by applying Lemma~\ref{lem:ptwofgc.1} and Theorem~\ref{thm:pliablemain},
we find a solution to a $\ptwofgc$ instance of cost $\leq\pliableapx\opt$.
Moreover, for even $p$, the approximation ratio is $6~(=4+2)$ since
the $\ptwofgc$ instance corresponds to an $\fcproblem$ such that
$f$ is an uncrossable function (by Lemma~\ref{lem:ptwofgc.1}),
hence, we find a solution of cost $\leq{2}\opt$ (by Theorem~\ref{thm:primal-dual-wgmv}).
This completes the proof of Theorem~\ref{thm:ptwofgc}.
\qedhere
\end{proofof}
}

\section*{Funding}
{
{The first author is partially supported by Office of Naval Research (ONR) Grant N00014-21-1-2575.}
{The second author is supported in part by NSERC, RGPIN-2019-04197.}
{The fourth author received funding from the following sources: NSERC grant 327620-09 and an NSERC DAS Award, European Research Council (ERC) under the European Union’s Horizon 2020 research and innovation programme (grant agreement no. ScaleOpt–757481), and Dutch Research Council NWO Vidi Grant 016.Vidi.189.087.}
}


\bigskip
\bigskip

\section*{Conflicts of interest / Competing interests}

{
The authors have no conflicts of interest to declare that are
relevant to the content of this article.


}

\bigskip
\bigskip

\section*{Acknowledgments}
We thank the anonymous reviewers and the ICALP PC for their comments.
We are grateful to Cedric Koh and Madison Van Dyk for reading a
preliminary version, and for their detailed comments and feedback.

{A preliminary version of this paper
appeared in the Proceedings of the
{50th International Colloquium on Automata, Languages, and Programming (ICALP 2023)},
Ed.~{K.Etessami, U.Feige, and G.Puppis}
\url{https://drops.dagstuhl.de/opus/volltexte/2023/18067/}
(LIPIcs, Volume 261, Article No.~15, pp.~15:1--15:19),
\cite{BCGI23}.
}

\bigskip
\bigskip
{
\bibliographystyle{sn-mathphys}
\bibliography{sn-bcgi-icalp-ref}

}

\appendix


\section{\GenUn families: an example} \label{sec:genun-family-example} \label{append:cylinders}
{
In Section~\ref{sec:general-WGMV}, we showed an approximation ratio
of $O(1)$ for the WGMV primal-dual method applied to any
$\fcproblem$ where $f$ is a \genun function satisfying property~\propred.
A natural question arises:
Is property~\propred\ essential for this result?
In other words, does the WGMV primal-dual method for the $\fcproblem$
where $f$ is a \genun function achieve an approximation ratio
of~$O(1)$?
The next construction shows that the answer is \textbf{no};
thus, property~\propred\ is essential for our results in
Section~\ref{sec:general-WGMV}.

Recall that a family of sets $\F\subseteq 2^V$ is called a {\genun family}
	if $A, B \in\F$ implies that at least two of the four sets
	$A\cup{B}, A\cap{B}, A\sm{B}, B\sm{A}$ also belong to $\F$.

\begin{figure}[htb]
	\centerline{\includegraphics[scale=0.4]{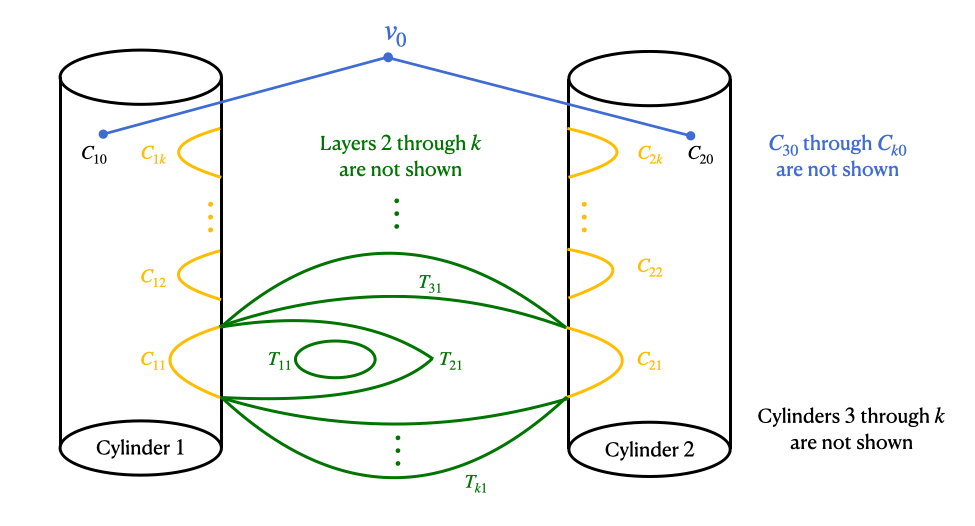}}
	\centerline{\includegraphics[scale=0.4]{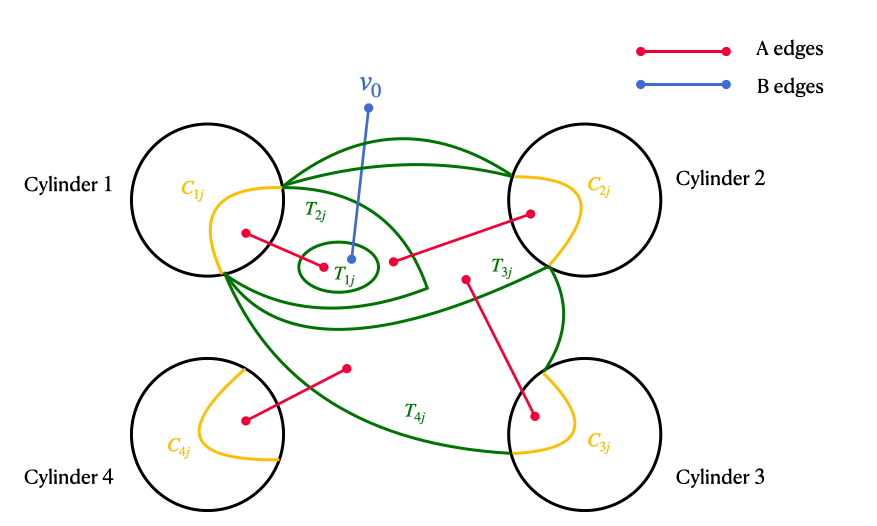}}
\caption{
	\label{fig:WGMV_counterexample}
	Top: A bad example for the primal-dual method for augmenting a \genun family.
	\newline
	Bottom: Edges of type~(A) and type~(B) in
	a bad example for the primal-dual method for augmenting a \genun family.}
\end{figure}

Now, we describe the construction: We have node-sets $C_{ij}$ for
$i=1,2,\ldots,k$ and $j=0,1,2,\ldots,k$. All these $C$-sets are
pair-wise disjoint. Additionally, for $j'=1,\ldots, k$, we have
node-sets $T_{1j'} \subsetneq T_{2j'}\subsetneq \cdots T_{kj'}$ and
each of these are disjoint from all the $C$-sets.  Additionally,
we have at least one node $v_0$ lying outside the union of all these
$C$-sets and $T$-sets. See Figure~\ref{fig:WGMV_counterexample}.
We designate $\mathcal{F'}$ to be the following family of node-sets
that consists of two types of sets:
\begin{align*}
& (I)  \qquad & C_i := \bigcup_{j=0}^k C_{ij} \quad               & \text{ for } i\in[k] \\
& (II) \qquad & T_{i'j'} \cup \bigcup_{(i,j) \in R'} C_{ij} \quad & \text{ for }
		i' \in[k],~ j'\in[k] \text{ and } \\
&&&		R' \subseteq
	{R(i')}
\text{ where } {R(i')} = \{(i,j) \:|\: 1\leq i < i',\; 0\leq j\leq k\}
\end{align*}
Informally speaking, the sets $C_1,C_2,\dots,C_k$ can be viewed as
(pairwise-disjoint) ``cylinders'', see Figure~\ref{fig:WGMV_counterexample},
the (first) index $i$ is associated with one of these cylinders, and
note that $C_{i0} = C_i \setminus (\cup_{j=1}^k C_{ij})$;
the (second) index $j$ is associated with a ``layer'' (i.e., a horizontal plane), and
the sub-family
	$T_{1j'} \subsetneq T_{2j'}\subsetneq \cdots T_{kj'}$
	forms a nested family on layer~$j'$, see Figure~\ref{fig:WGMV_counterexample}.
For notational convenience, let $T_{0j'}=\emptyset,\; \forall j'\in[k]$.
Observe that a set of type~(II) is the union of one set $T_{i'j'}$ of
the nested family of layer~$j'$
together with the sets of an arbitrary sub-family of each of the
``cylinder families'' $\{ C_{i0}, C_{i1}, C_{i2}, \dots, C_{ik} \}$
with (first) index
{$i < i'$}.
We claim that the family $\mathcal{F'}$ satisfies the condition of
Lemma~\ref{lem:genuncross}. Indeed consider $A,B \in \mathcal{F'}$.
If $A$ and $B$ are of type~(I), then they are disjoint. If $A$ is
of type~(I) and $B$ is of type~(II) such that
{$A\cap{B}\not=\emptyset$},
then $A\cup B$ and $B\setminus{A}$ are sets of type~(II) so both belong
to $\mathcal{F'}$. If $A$ and $B$ are of type~(II) such that both
have the same (second) index $j'$, then both $A\cup{B}$ and $A\cap{B}$
are sets of type~(II).  On the other hand, if $A$ and $B$ are of
type~(II) such that the (second) index $j'$ is different for $A,B$,
then $A\setminus{B}$ and $B\setminus{A}$ are sets of type~(II).
Thus,
{by Lemma~\ref{lem:genuncross},}
adding in all the complements of the sets of this family gives
us a \genun family $\mathcal{F}$.

The edges of the graph are of two types:

\begin{enumerate}
\item[(A)] For $i' = 1,\ldots,k$ and $j' = 1,\ldots, k$, we have
an edge from $T_{i'j'} \setminus T_{(i'-1)j'}$ to $C_{ij}$ where $i=i'$ and $j=j'$.

\item[(B)] For $j=1,\ldots, k$, we have an edge from $v_0$ to $T_{1j}$.
For $i=1,\ldots,k$, we have an edge from $v_0$ to $C_{i0}$.
\end{enumerate}

When the primal-dual algorithm is applied to this instance, then
it picks all the edges of type~(A); note that there are $k^2$ edges of type~(A).
Let us sketch the working of the primal-dual algorithm on this instance.
{
\begin{itemize}
\item 
Initially, the active sets are the type~(I) sets $C_1,\dots,C_k$
and the smallest $T$-sets $T_{11},\dots,T_{1k}$. The algorithm
increases the dual variables of each of these sets by $1/2$ and
then picks all the type~(A) edges between $T_{1j}$ and $C_{1j}$ for $j\in[k]$.

\item 
In the next iteration, the (new) active sets are the type~(I) sets
$C_2,\dots,C_k$ and the type~(II) sets of the form $T_{2j} \cup
C_{1j}$ for $j\in[k]$. The algorithm increases the dual variables
of each of these sets by $1/4$ and then picks all the type~(A) edges
between $T_{2j}$ and $C_{2j}$ for $j\in[k]$.

\item 
Similarly, in the $i^{th}$ iteration, the active sets are the
type~(I) sets $C_i,\dots,C_k$ and the type~(II) sets of the form $T_{ij}
\cup C_{1j} \cup\dots\cup C_{(i-1)j}$ for $j\in[k]$. The algorithm
increases the dual variables of each of these sets by $1/2^i$ and then picks all
the type~(A) edges between $T_{ij}$ and $C_{ij}$ for $j\in[k]$.

\item 
Finally, the reverse-delete step does not delete any
{edge, because the edges of type~(A) form an inclusion-wise minimal solution}.

\end{itemize}
}

On the other hand, all the edges of type~(B) form a feasible solution
of this instance, and there are $\leq 2k$ such edges.  In more
detail, each of the type~(I) sets $C_i$ is covered by the type~(B)
edge between $v_0$ and $C_{i0}$, and each of the type~(II) sets
containing $T_{i'j'}$ is covered by the type~(B) edge between $v_0$
and $T_{1j'}$. Thus, the optimal solution picks $\leq 2k$ edges.

}
{
\section{Missing Proofs from Section~\ref{sec:general-WGMV}} \label{append:proofsec3}

This section has several lemmas and proofs from Section~\ref{sec:general-WGMV}
that are used to prove our main result, Theorem~\ref{thm:pliablemain}.

\begin{lemma}\label{lem:uncrossingwintesssets}
Suppose $S_1$ is a witness for edge $e_1$ and $S_2$ is a witness
for edge $e_2$ such that $S_1$ overlaps $ S_2$. Then there exist
$S_1'$ and $S_2'$ satisfying the following properties:

\begin{enumerate}
    \item [(i)]$S_1'$ is a valid witness for edge $e_1$, $S_2'$
    is a valid witness for edge $e_2$, and $S_1'$ does not overlap $S_2'$.

    \item[(ii)]$S_1',S_2' \in \{S_1,S_2,S_1\cup S_2, S_1\cap S_2,
    S_1\setminus S_2, S_2\setminus S_1\}$.

    \item[(iii)] either $S_1' = S_1$ or $S_2' = S_2$.
\end{enumerate}
\end{lemma}

\begin{proof}
{
{We prove the lemma via an exhaustive case analysis.}
Note that at least two of the four sets $S_1\cup S_2, S_1\cap
S_2, S_1\setminus S_2, S_2\setminus S_1$ must be violated in the
current iteration.
{
Moreover, observe that $e_1\in\delta_E(S_1\sm{S_2})$ or $e_1\in\delta_E(S_1\cap{S_2})$,
and in the latter case
$e_1\in E(S_1\cap{S_2}, V\sm(S_1\cup{S_2}))$ or
$e_1\in E(S_1\cap{S_2}, S_2\sm{S_1})$.}
We consider the following cases.

\begin{enumerate}
    \item If $S_1 \cup S_2$ and $S_1 \cap S_2$ are violated or
    $S_1\setminus S_2$ and $S_2\setminus S_1$ are violated, then
    the proof of Lemma 5.2 in \cite{WGMV95} can be applied.

    \item
	{Suppose $S_1 \cap S_2$ is violated, and one of
	$S_1\sm{S_2}$ or $S_2\sm{S_1}$ is violated. W.l.o.g.\ suppose
	$S_1 \cap S_2$  and $S_1\sm{S_2}$ are violated (the other
	case is similar).
	Consider where the end-nodes of the edge $e_1$ lie.
	If $e_1\in\delta_E(S_1\sm{S_2})$, then fix $S_1':=S_1\sm{S_2}$
	and $S_2':=S_2$;
	otherwise, $e_1\in\delta_E(S_1\cap{S_2})$, and then fix $S_1':=S_1\cap{S_2}$
	and $S_2':=S_2$.}

    \item
	{Suppose $S_1 \cup S_2$ is violated, and one of
	$S_1\sm{S_2}$ or $S_2\sm{S_1}$ is violated. W.l.o.g.\ suppose
	$S_1 \cup S_2$  and $S_1\sm{S_2}$ are violated (the other
	case is similar).
	Consider where the end-nodes of the edges $e_1$ and $e_2$ lie.
	If $e_1\in\delta_E(S_1\sm{S_2})$, then fix $S_1':=S_1\sm{S_2}$ and $S_2':=S_2$;
	similarly, if $e_1\in E(S_1\cap{S_2}, V\sm(S_1\cup{S_2}))$,
	then fix $S_1':=S_1\cup{S_2}$ and $S_2':=S_2$;
	finally, if $e_1\in E(S_1\cap{S_2}, S_2\sm{S_1})$,
	then fix $S_1':=S_1$, and then
	if $e_2\in\delta_E(S_1\cup{S_2})$, then fix $S_2':=S_1\cup{S_2}$,
	otherwise, fix $S_2':=S_1\sm{S_2}$.}
\end{enumerate}
This completes the proof of the lemma.
}
\end{proof}

\begin{lemma}\label{lem:nonoverlappreserved}
    Suppose a set $A_1$ overlaps a set $A_2$ and a third set $A_3$
    does not overlap $A_1$ nor $A_2$.
    Then $A_3$ does not overlap any of the sets $A_1\cup A_2, A_1\cap
    A_2, A_1 \setminus A_2, A_2\setminus A_1$.
\end{lemma}

\begin{proof}
    Note that since $A_3$ does not overlap $A_1$ (or $A_2$), they
    are either disjoint or one contains the other. We consider the following cases.

    \begin{enumerate}
	\item Suppose $A_3 \cap A_1 = \emptyset$. Then \myredtext{$A_2\not\subseteq
	A_3$} since $A_1 \cap A_2 \neq \emptyset$. If $A_3 \cap A_2
	= \emptyset$, then $A_3 \subseteq V\setminus A_1\cup A_2$
	and we are done. Finally if $A_3 \subseteq A_2$, then $A_3
	\subseteq A_2\setminus A_1$ and we are done.

	\item Suppose $A_1 \subseteq A_3$. Then $A_3 \cap A_2 \neq
	\emptyset$ since $A_1\cap A_2 \neq \emptyset$. Also,
	\myredtext{$A_3\not\subseteq A_2$ since $A_1\not\subseteq A_2$}. If $A_2
	\subseteq A_3$, then $A_1\cup A_2 \subseteq A_3$ and we are done.

	\item Suppose $A_3 \subseteq A_1$. Then \myredtext{$A_2\not\subseteq A_3$} since
	$A_2\sm{A_1}\neq\emptyset$. If $A_3 \subseteq A_2$, then $A_3\subseteq
	A_1\cap A_2$ and we are done. Finally if $A_3\cap A_2 =
	\emptyset$, then $A_3 \subseteq A_1 \setminus A_2$ and we are done.
    \end{enumerate}
\end{proof}

\noindent
\textbf{Lemma~\ref{lem:lamfamily}. }
\textit{
There exists a laminar family of witness sets.
}

\begin{proof}
{We show that any witness family can be transformed to a laminar family by
repeatedly applying Lemma~\ref{lem:uncrossingwintesssets}.}
We prove this by induction on the size of the witness family $\ell$.

\medskip

\noindent
\textbf{Base Case:} Suppose $\ell = 2$, then one application of
Lemma~\ref{lem:uncrossingwintesssets} is sufficient.

\medskip

\noindent
\textbf{Inductive Hypothesis:} If $S_1,\ldots, S_\ell$ are witness
sets for edges $e_1,\ldots, e_\ell$ respectively with $\ell \leq
k$, then, by repeatedly applying Lemma~\ref{lem:uncrossingwintesssets},
one can construct witness sets $S'_1,\ldots, S'_\ell$ for the edges
$e_1,\ldots, e_\ell$ respectively such that $S'_1,\ldots, S'_\ell$
is a laminar family.

\medskip

\noindent
\textbf{Inductive Step:} Consider $k+1$ witness sets $S_1,\ldots,
S_{k+1}$.
{By the inductive hypothesis, we can repeatedly apply
Lemma~\ref{lem:uncrossingwintesssets} to all the witness sets
$S_1,\ldots, S_k$ and obtain witness sets $S_1',\ldots, S_k'$ that
form a laminar family.}
We now consider the following cases.

\begin{enumerate}
    \item If $S_{k+1}$ does not overlap some $S_i'$, say $S_1'$,
    then we can apply the inductive hypothesis to the $k$ sets
    $S_2',\ldots, S_k', S_{k+1}$ and we obtain a laminar family
    of witness sets, none of which overlap $S_1'$ either (by
    Lemma~\ref{lem:nonoverlappreserved}) and so we are done.

    \item Suppose $S_{k+1}$ overlaps all the sets $S_1',\ldots, S_k'$
    and for some $S_i'$, say $S_1'$, applying 
    Lemma~\ref{lem:uncrossingwintesssets} to the pair $S_1',S_{k+1}$ gives
    $S_1', S_{k+1}'$. Then $S_1'$ does not overlap any of the witness
    sets $S_2',\ldots, S_{k+1}'$, hence, applying the inductive
    hypothesis to these $k$ sets gives us a
    laminar family of witness sets $S_2^{''},\ldots, S_k^{''}$.
    By Lemma~\ref{lem:nonoverlappreserved}, $S_1'$ does not overlap
    any of the sets $S_2^{''},\ldots, S_k^{''}$ and so we are done.

    \item Suppose $S_{k+1}$ overlaps all the sets $S_1',\ldots, S_k'$
    and, for every $S_i'$, applying Lemma~\ref{lem:uncrossingwintesssets}
    to the pair $S_i', S_{k+1}$ gives $S''_i, S_{k+1}$. Then after
    doing this for every $S_i'$, we end up with the witness
    family $S_1^{''}, \ldots, S_k^{''}, S_{k+1}$ with the property
    that $S_{k+1}$ does not overlap any of the other sets. Applying
    the inductive hypothesis to the $k$ sets $S_1^{''}, \ldots, S_k^{''}$ 
    gives us a laminar family of witness sets $S_1^{'''},\ldots, S_k^{'''}$.
    By Lemma~\ref{lem:nonoverlappreserved}, $S_{k+1}$ does not overlap
    any of the sets $S_1^{'''},\ldots, S_k^{'''}$ and so we are done. \qedhere 
\end{enumerate}
\end{proof}

}
{
\section{Proof of Lemma~\ref{lem:ptwofgc.1} from Section~\ref{sec:p2fgc}} \label{append:proofsec6}

This section has a proof of Lemma~\ref{lem:ptwofgc.1}.
This proof is the same as the proof we posted on Arxiv in 2022, \cite{BCGI:arxiv}.

\noindent
\textbf{Lemma~\ref{lem:ptwofgc.1}. }
\textit{
$f$ is a \genun function that satisfies property~\propred. Moreover, for even $p$,  $f$ is an uncrossable function. 
}

{
\begin{proof}
Consider two violated sets $A, B \subsetneq V$. W.l.o.g.\ we may
assume that both $A$ and $B$ contain a fixed node $r \in V$.
If $A,B$ do not cross, then it is easily seen that 
$f(A)+f(B) = \max( f(A\cap{B})+f(A\cup{B}),\; f(A\sm{B})+f(B\sm{A}) )$,
hence, the inequality for \genun functions holds for $A,B$.
Thus, we may assume that $A,B$ cross.
The following equations hold, see Frank's book \cite[Chapter~1.2]{Frank2011}:

\begin{align}
& |\delta(A)| = |\delta(B)| = p+1 \label{eq1} \\
& |\delta(A \cup B)| + |\delta(A \cap B)| + 2|F(A \setminus B, B \setminus A)| = |\delta(A)| + |\delta(B)|  \label{eq2} \\
& |\delta(A \setminus B)| + |\delta(B \setminus A)| + 2|F(A \cap B, V \setminus (A \cup B))| = |\delta(A)| + | \delta(B)| \label{eq3} \\
& |\delta(A \setminus B)| + |\delta(A \cap B)| =  |\delta(A)| + 2 |F(A \setminus B, A \cap B)| \label{eq4}
\end{align}

Since $|\delta(S)|\geq{p}, \forall\emptyset\neq{S}\subsetneq{V}$,
equations~\eqref{eq1},~\eqref{eq2}~and~\eqref{eq3} imply that
$|\delta(A \cup B)|, |\delta(A \cap  B)|,  |\delta(A \setminus B)|,
|\delta(B \setminus A)| \in \{p,p+1,p+2\}$, and, moreover,
  $|F(A \sm B, B \sm A)| \leq{1},\: |F(A \cap B, {V\sm(A \cup B)})| \leq{1}$.

Furthermore, the above four equations imply the following parity-equations
(equations~\eqref{eq5},~\eqref{eq6}~and~\eqref{eq7} follow from
equations~\eqref{eq2},~\eqref{eq3}~and~\eqref{eq4}, respectively).

\begin{align}
& |\delta(A \cup B)| \equiv |\delta(A \cap B)| \quad  (\text{mod} \, 2) & \label{eq5} \\
& |\delta(A \sm B)| \equiv |\delta(B \sm A)| \quad  (\text{mod} \, 2) & \label{eq6} \\
& |\delta(A\cap B)| \equiv |\delta(A\setminus B)| + |\delta(A)| \equiv
	|\delta(B\setminus A)| + |\delta(B)| \quad  (\text{mod} \, 2) & \label{eq7}
\end{align}

\medskip
\noindent
\textbf{Case 1} (of proof of the lemma):
Suppose that $p$ is even.
Then the parity-equations \eqref{eq5}--\eqref{eq7} imply that among the two pairs of
cuts $\{\delta(A\cup{B}), \delta(A\cap{B})\}$ and $\{\delta(A\sm{B}),
\delta(B\sm{A})\}$, one of the pairs consists of two $(p+1)$-cuts,
and the other pair has at least one cut of size $p$.
Since $F$ is a feasible solution of $\genfgc{p}{1}$,
every $p$-cut of $(V,F)$ consists of safe edges (that is, a $p$-cut
cannot contain any unsafe edge).
Since $A$ and $B$ are violated sets, each of the cuts $\delta(A)$
and $\delta(B)$ contains at least two unsafe edges.

\begin{claim} \label{clm:ptwofgc.1}
Each cut $\delta(S)$ of the pair of $(p+1)$-cuts
{$\{\delta(A\cup{B}), \delta(A\cap{B})\}$ or
$\{\delta(A\sm{B}), \delta(B\sm{A})\}$ (obtained from $A,B$)}
also contains at least two unsafe edges, and so $S$ is a violated set
for each of these cuts.
\end{claim}

We prove this claim by a simple case analysis on the end-nodes of
the unsafe edges of the cuts $\delta(A)$ and $\delta(B)$.
W.l.o.g.\ assume that $\delta(A\cup{B})$ and $\delta(A\cap{B})$ are
$(p+1)$-cuts and that $\delta(A\setminus{B})$ is a $p$-cut. The
other cases are handled similarly. Clearly, $\delta (A\setminus{B})$
has no unsafe edges.
Since $A$ is a violated set, either (i)~$F(A\cap B, B\sm{A})$ has two unsafe edges or (ii)~$F(A\cap B, B\sm{A})$ has one
unsafe edge and $F(A\cap B, {V\sm(A\cup B)})$ has one unsafe
edge (recall that the size of the latter edge-set is $\leq1$).
Since $B$ is a violated set, either
(iii)~$F(B\sm{A}, {V\sm(A\cup{B})})$ has two unsafe edges
or (iv)~$F(B\sm{A}, {V\sm(A\cup B)})$ has one unsafe edge and
$F(A\cap B, {V\sm(A\cup B)})$ has one unsafe edge.
Now, observe that both $A\cap{B}$ and $A\cup{B}$ are violated sets, because,
by (i)~or~(ii), $\delta(A\cap{B})$ has at least two unsafe edges, and,
by (iii)~or~(iv), $\delta(A\cup{B})$ has at least two unsafe edges.
$\big($If both $\delta(A\sm{B})$ and $\delta(B\sm{A})$ are $(p+1)$-cuts, and
there are no unsafe edges in either $\delta(A\cap{B})$ or $\delta(A\cup{B})$,
then both $\delta(A\sm{B})$ and $\delta(B\sm{A})$
have $\geq2$ unsafe edges, so both $A\sm{B}$ and $B\sm{A}$ are violated sets.$\big)$
This proves the claim.

Therefore, when $p$ is even, the function $f$ is an uncrossable
function (recall that every uncrossable function is a \genun function that satisfies property~\propred).

\medskip
\noindent
\textbf{Case 2} (of proof of the lemma):
Suppose that $p$ is odd.
Then we have
$ |\delta(A \cup B)| \equiv |\delta(A \cap B)| \equiv
	|\delta(A\sm{B})| \equiv |\delta(B\sm{A})| \quad  (\text{mod} \, 2)$.
Hence, the above equations \eqref{eq1}--\eqref{eq4} and parity-equations
\eqref{eq5}--\eqref{eq7} imply that
either all four cuts
$\delta(A\cup{B}), \delta(A\cap{B}), \delta(A\sm{B}), \delta(B\sm{A})$
are $(p+1)$-cuts, or at least one
cut from each pair $\{\delta(A\cup{B}), \delta(A\cap{B})\}$ and
$\{\delta(A\sm{B}), \delta(B\sm{A})\}$ is a $p$-cut.

\begin{claim} \label{clm:ptwofgc.2}
Suppose that at least one cut from each pair $\{\delta(A\cup{B}),
\delta(A\cap{B})\}$ and $\{\delta(A\sm{B}), \delta(B\sm{A})\}$ is
a $p$-cut.
Then either $A$ is not a violated set, or $B$ is not a violated set.
\end{claim}

To prove this claim, let us assume that $|\delta(A\sm{B})|=p$;
similar arguments apply for the other case.
The edge-set $F$ is feasible for $\genfgc{p}{1}$, hence, all edges
of $\delta(A\sm{B})$ are safe.  Moreover, one of the cuts
$\delta(A\cap{B})$ or $\delta(A\cup{B})$ has size $p$ and consists
of safe edges.
If $|\delta(A\cap{B})|=p$, then the cut $\delta(A)$ consists of safe edges
(since $\delta(A)\subseteq\delta(A\cap{B})\cup\delta(A\sm{B})$),
hence, $A$ cannot be a violated set.
Otherwise, if $|\delta(A\cup{B})|=p$, then the cut $\delta(B)$ consists of safe edges
(since $\delta(B)=\delta(V\sm{B})\subseteq\delta(A\sm{B})\cup\delta(V\sm(A\cup{B}))$),
hence, $B$ cannot be a violated set.
This proves the claim.

Claim~\ref{clm:ptwofgc.2} implies that all four cuts
$\delta(A\cup{B}), \delta(A\cap{B}), \delta(A\sm{B}), \delta(B\sm{A})$
are $(p+1)$-cuts.

\begin{claim} \label{clm:ptwofgc.3}
Consider two crossing violated sets $A, B \subseteq V$.
\begin{enumerate}
\item[(i)]
Then each of the four cuts
$\delta(A\cup{B}), \delta(A\cap{B}), \delta(A\sm{B}), \delta(B\sm{A})$
has size $(p+1)$, and we have
$|F(A\sm{B}, B\sm{A})| = 0 = |F(A\cap{B}, V\sm(A\cup{B}))|$.

\item[(ii)]
Moreover, each of the four sets
$F(A\cap{B},\,A\sm{B}),\; F(A\cap{B},\,B\sm{A}),\;
	F(A\sm{B},\,V\sm(A\cup{B})),\; F(B\sm{A},\,V\sm(A\cup{B}))$
has size $(p+1)/2$.
\end{enumerate}
\end{claim}

Part~(i) of this claim follows from the above arguments
(see the discussion before and after Claim~\ref{clm:ptwofgc.2}).
Moreover, we have
$|F(A \setminus B, B \setminus A)| = 0$ and
$|F(A \cap B, V \setminus (A \cup B))| = 0$ by equations~\eqref{eq2},\eqref{eq3}.
Next, consider part~(ii) of the claim.
Using the first part of the claim together with equation~\eqref{eq4},
we have $|F(A\cap{B},\,A\sm{B})|=(p+1)/2$.
Since $\delta(A\sm{B})$ is a $(p+1)$-cut and it is the disjoint union of
$F(A\sm{B},\,A\cap{B})$ and $F(A\sm{B},\,V\sm(A\cup{B}))$,
we have $|F(A\sm{B},\,V\sm(A\cup{B}))|=(p+1)/2$.
Similarly, we have $|F(A\cap{B},\,B\sm{A})|=(p+1)/2$, $|F(B\sm{A},\,V\sm(A\cup{B}))|=(p+1)/2$.
This proves the claim.

\begin{claim} \label{clm:ptwofgc.4}
\begin{enumerate}
\item[(i)]
At least two of the four cuts
$\delta(A\cup{B}), \delta(A\cap{B}), \delta(A\sm{B}), \delta(B\sm{A})$
each contain at least two unsafe edges.

\item[(ii)]
Moreover, if $A$ is a minimal violated set,
then $F(B\sm{A},\,V\sm(A\cup{B}))$ has at least two unsafe edges, and
each of $F(A\cap{B},\,B\sm{A})$ and $F(A\sm{B},\,V\sm(A\cup{B}))$ has exactly one unsafe edge.
\end{enumerate}
\end{claim}

We prove this claim by a simple case analysis on the end-nodes of
the unsafe edges of the cuts $\delta(A)$ and $\delta(B)$.
Observe that
$|F(A \setminus B, B \setminus A)| = 0$ and
$|F(A \cap B, V \setminus (A \cup B))| = 0$,
by equations~\eqref{eq2},\eqref{eq3}.
Each edge of $\delta(A)$ is in exactly one of the sets
$F(A\cap{B},\,B\sm{A})$ or $F(A\sm{B},\,V\sm(A\cup{B}))$;
call these edge-sets $\phi_1, \phi_2$ for notational convenience.
Similarly, each edge of $\delta(B)$ is in exactly one of the sets
$F(A\cap{B},\,A\sm{B})$ or $F(B\sm{A},\,V\sm(A\cup{B}))$;
call these edge-sets $\phi_3, \phi_4$ for notational convenience.
If two of the unsafe edges of $\delta(A)$ are in the same set
(i.e., if $\phi_1$ or $\phi_2$ has two unsafe edges),
then part~(i) of the claim holds, since either
$\delta(A\cap{B}), \delta(B\sm{A})$ each have two (or more) unsafe edges or else
$\delta(A\cup{B}), \delta(A\sm{B})$ each have two (or more) unsafe edges.
Similarly, if two of the unsafe edges of $\delta(B)$ are in the same set
(i.e., if $\phi_3$ or $\phi_4$ has two unsafe edges),
then part~(i) of the claim holds.
There is one remaining case: each of the four sets
$F(A\cap{B},\,B\sm{A})$, $F(A\sm{B},\,V\sm(A\cup{B}))$,
$F(A\cap{B},\,A\sm{B})$, $F(B\sm{A},\,V\sm(A\cup{B}))$
has exactly one unsafe edge.
In this case, each of the four sets
$\delta(A\cup{B}), \delta(A\cap{B}), \delta(A\sm{B}), \delta(B\sm{A})$
has two unsafe edges. This proves part~(i) of the claim.
To prove part~(ii) of the claim, observe that
neither of the $(p+1)$-cuts $\delta(A\cap{B})$ or $\delta(A\sm{B})$
can have two (or more) unsafe edges,
otherwise, a proper subset of the minimal violated set $A$ would be violated.
Then, the above case analysis shows that each of the two sets
$F(A\cap{B},\,B\sm{A})$, $F(A\sm{B},\,V\sm(A\cup{B}))$
has exactly one unsafe edge, and the set
$F(B\sm{A},\,V\sm(A\cup{B}))$ has two (or more) unsafe edges.
This proves part~(ii) of the claim.

Clearly, the function $f$ is a \genun function, by Claim~\ref{clm:ptwofgc.4}, part~(i).
Next, we argue that the function $f$ satisfies property~\propred.

Let $C$ be a minimal violated set and let
$S_1,S_2$ be violated sets such that $C$ crosses
both $S_1,S_2$ and $S_1 \subsetneq S_2$. 
W.l.o.g.\ assume that $S_2 \sm (S_1\cup C)$ is non-empty. 
Since $C$ crosses $S_2$, the edge-set
$F(S_2\sm{C},\,V\sm(S_2\cup{C}))$
has size $(p+1)/2$, by Claim~\ref{clm:ptwofgc.3}.
Moreover, by Claim~\ref{clm:ptwofgc.4}, part~(ii), this edge-set contains two unsafe edges.
Observe that $S_1\cup{C}$ crosses $S_2$; to see this, note that
$C$ crosses $S_2$ so the sets $C\cap{S_2},\, C\sm{S_2},\, V\sm(C\cup{S_2})$
are non-empty, and, by assumption, $S_2 \sm (S_1\cup C)$ is non-empty.
Applying Claim~\ref{clm:ptwofgc.3} to this pair of crossing sets, we see that the edge-set
$F(S_2\sm(S_1\cup{C}),\,V\sm(S_2\cup{C}))$
has size $(p+1)/2$.
Then we have
$F(S_2\sm{C},\,V\sm(S_2\cup{C})) = F(S_2\sm(S_1\cup{C}),\,V\sm(S_2\cup{C}))$,
because both edge-sets have the same size and one edge-set is a subset of the other edge-set.
Hence, $F(S_2\sm(S_1\cup{C}),\,V\sm(S_2\cup{C}))$ contains two unsafe edges.
Finally, by Claim~\ref{clm:ptwofgc.3}, the cut
$\delta(S_2\sm(S_1\cup{C}))$
has size $(p+1)$.
Since this cut has two (or more) unsafe edges, $S_2\sm(S_1\cup{C})$ is a violated set.
This proves that the function $f$ satisfies property~\propred.

This completes the proof of Lemma~\ref{lem:ptwofgc.1}.
\end{proof}
}
}
{
\section{Optimal Dual Solutions with Non-Laminar Supports} \label{append:nonlaminar}

In this section, we describe an instance of the $\augsmallcuts$ problem where none of the optimal dual solutions (to the dual LP
given in \eqref{eq:primaldualLPs}, Section~\ref{sec:prelims}) have a laminar support.  
Recall that the connectivity requirement function $f$ for the
$\augsmallcuts$ problem is \genun and satisfies property~\propred,
as seen in the proof of Theorem~\ref{thm:nearmincutsaugmentation}.

Consider the graph $G = (V,E)$ (shown in Figure~\ref{fig:nonlaminar} below using solid edges) which is a cycle on $4$ nodes $1, 2, 3, 4$, in that order.
Edge-capacities are given by $u_{12}=3, u_{23}=4, u_{34}=2, u_{41}=1$.
The link-set (shown using dashed edges) is $L=\{ 12, 23, 34, 41 \}$, which is
a disjoint copy of $E$. Link-costs are given by
$c_{12} = c_{23} = c_{34} = 1$ and $c_{41} = 2$.

\begin{figure}[ht]
    \centering
    \includegraphics[width=0.5\textwidth]{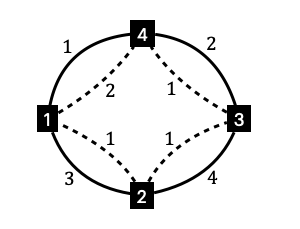}
    \vspace{-10pt}
    \caption{An instance of the $\augsmallcuts$ problem where every optimal dual solution has non-laminar support.}
    \label{fig:nonlaminar}
\end{figure}

Consider the $\augsmallcuts$ instance that arises when we choose
$\capbound = 6$. The family of small cuts (with capacity strictly
less than $\capbound$) is given by $\bigcup_{S \in \As} \{S, V
\setminus S\}$, where
\[
\As = \{ \{1\}, \{1,2\}, \{2,3\}, \{1,2,3\} \}.
\]
The associated \genun function $f$ satisfies $f(S) = 1$
if and only if $S\in\As$ or $V\setminus{S}\in\As$ holds. 
Observe that $f$ is \emph{not} uncrossable since $f(\{1,2\}) = 1 = f(\{2,3\})$, but
$f(\{1,2\} \cap \{2,3\}) = f(\{2\}) = 0$ and $f(\{2,3\} \sm \{1,2\}) = f(\{3\}) =0$.
Also note that the minimal violated set $\{2,3\}$ (w.r.t. $F = \emptyset$) crosses the violated set $\{1,2\}$.

It can be seen that there are three inclusion-wise minimal link-sets that are feasible for the above instance and these are given by
\newcommand{\feasLinks}{\ensuremath{\mathcal{C}}}
\[
\feasLinks := \{ \{12,23,34\}, \{12,41\}, \{34,41\} \}.
\]
Since each $F \in \feasLinks$ has cost~$3$, the optimal value for the instance is~$3$. 
Next, since $L$ contains at least two links from every nontrivial cut,
the vector $x\in[0,1]^L$ with $x_e=\frac12,\,\forall{e}\in{L}$
is a feasible augmentation for the fractional version of the instance,
i.e., $x$ is feasible for the primal LP given in
\eqref{eq:primaldualLPs}, Section~\ref{sec:prelims}.
Therefore, the optimal value of the primal LP is at most $\frac52$.

Now, consider the dual LP, which is explicitly stated below. The
dual packing-constraints are listed according to the following
ordering of the links: $12, 23, 34, 41$.
For notational convenience, we use the shorthand $y_1$ to denote
the dual variable $y_{\{1\}}$ corresponding to the set $\{1\}$.
We use similar shorthand to refer to the dual variables of the other sets;
thus, $y_{234}$ refers to the dual variable $y_{\{2,3,4\}}$, etc.
\begin{minipage}{\textwidth}
\begin{align*}
& \max &  (y_{1} + y_{234}) & \quad + & (y_{12} + y_{34}) & \quad + & (y_{23} + y_{14}) & \quad + & (y_{123} + y_{4}) & \qquad &  \\
& \, \text{subject to:} & (y_{1} + y_{234}) & & & \quad + &  (y_{23} + y_{14}) & & & \; \leq 1 \; \\
& & & & (y_{12} + y_{34}) & & & & & \; \leq 1 \; \\
& & & & & & (y_{23} + y_{14}) & \quad + & (y_{123} + y_{4}) & \; \leq 1 \; \\
& & (y_{1} + y_{234}) & \quad + & (y_{12} + y_{34}) & & & \quad + & (y_{123} + y_{4}) & \; \leq 2 \; \\
& & & & & & & & y & \; \geq 0.
\end{align*}
\end{minipage}
Observe that adding all packing constraints gives $2 \cdot
\sum_{S\in\As} (y_S + y_{V\sm{S}}) \leq 5$, hence, the optimal value of the dual LP is at most $5/2$. Moreover, a feasible dual solution with objective $5/2$ must satisfy the following conditions:
\[
y_{1} + y_{234} = y_{23} + y_{14} = y_{123} + y_{4} = \frac12 \quad \text{ and } \quad y_{12} + y_{34} = 1.
\]
Clearly, there is at least one solution to the above set of equations, hence, by LP~duality, the optimal value of both the primal~LP and the dual~LP is $5/2$.

Furthermore, any optimal dual solution $y^*$ satisfies
 $\max(y^*_S,y^*_{V\sm{S}}) > 0$ for all $S \in \As$ (by the above set of equations).
We conclude by arguing that for any optimal dual
solution $y^*$, its support $\Ss(y^*) = \{ S \subseteq V : y^*_S > 0 \}$
is non-laminar, because some two sets $A, B \in \Ss(y^*)$ cross.
Since the relation $A$ crosses $B$ is closed under taking set-complements
(w.r.t.\ the ground-set $V$), we may assume w.l.o.g.\
that the support contains each set in
$\As = \{ \{1\}, \{1,2\},  \{2,3\}, \{1,2,3\} \}$.
The support of $y^*$ is not laminar because $\{1,2\}$ and $\{2,3\}$ cross.
}

\end{document}